\documentclass[11pt,letterpaper]{article}
\usepackage{amsmath,amssymb}
\usepackage{amsmath,amssymb,epsfig,cite}
\usepackage{xcolor,tikz-cd}
\usepackage{multirow,longtable}

\newcommand{\preprintnumber}[1]{\vbox{ \baselineskip 14pt \hfill
    \hbox{\normalsize UCI-TR-2019-03 } \\
\hfill \hbox{\normalsize #1} } \vskip 2cm}
\newcommand{\email}[1]{\footnote{email:#1}}

\begin{document}

\newcommand{\be}{\begin{equation}}
\newcommand{\ee}{\end{equation}}
\newcommand{\C}{{\mathbb C}}
\newcommand{\m}{{\rm m}}
\newcommand{\N}{{\rm N}}
\renewcommand{\P}{{\mathbb P}}
\newcommand{\R}{{\mathbb R}}
\newcommand{\tr}{{\rm tr}}
\newcommand{\T}{{\rm T}}
\newcommand{\Tr}{{\rm Tr}}
\newcommand{\U}{{\rm U}}

\newcommand{\Z}{{\mathbb Z}}
\newcommand{\blue}{\color{blue}}

\title{\preprintnumber{ EPHOU-19-001 }
Transitions of Orbifold Vacua}
\author{Kang-Sin Choi$^{\,a,b}$\email{kangsin@ewha.ac.kr} \ and Tatsuo Kobayashi$^{\,c}$\email{kobayashi@particle.sci.hokudai.ac.jp} \\
\it \normalsize $^a$ Scranton Honors Program, Ewha Womans University, Seoul 03760, Korea \\
\it \normalsize $^b$ Department of Physics and Astronomy, University of California, Irvine, California 92697, USA \\
\it \normalsize $^c$ Department of Physics, Hokkaido University, Sapporo 060-0810, Japan}
\date{}
\maketitle

\begin{abstract}
We study the global structure of vacua of heterotic strings compactified on orbifolds $T^4/\Z_N \, (N=2,3)$ in the presence of heterotic 5-branes. Gauge symmetry breaking associated with orbifold is described by instantons in the field theory. Phase transition between small instantons and heterotic 5-branes provides top-down, stringy account to the spectrum and modular invariance condition. Also it takes us from one vacuum to another by emitting and absorbing instantons. This means that many vacua with different gauge theory are in fact connected and are inherited from perturbative vacua. It follows that there are also transitions among twisted fields, heterotic 5-branes and instantons.
\end{abstract}

\newpage
%\tableofcontents

\section{Introduction}

Well-known is phase transition of small instantons, by which instantons may shrink to zero size and become heterotic 5-branes  \cite{Witten:1995gx,Duff:1996rs,Ganor:1996mu}. It recently attracted renewed attention in various contexts. For instance, geometry of F-theory helps us  identify non-perturbative states and describe the phase transitions \cite{MV,Bershadsky:1996nh,Haghighat:2014pva,Kim:2014dza,Choi:2017vtd,Choi:2017luj}. Such transition is important in understanding six dimensional superconformal field theory (SCFT) which is based on dynamics of M5-branes, dual to these heterotic 5-branes. In particular, the mysterious non-Abelian nature of such SCFT is associated with the dynamics of M5-branes. It is also tightly associated to the dynamics of non-critical strings, describing fluctuation of M5-branes, which can be studied by string theory.

In this paper, we use this phase transition to study vacua of heterotic strings, $E_8\times E_8$ and $SO(32)$, compactified on toroidal orbifolds $T^4/\Z_N, N=2,3$ \cite{DHVW,Erler:1993zy,Kachru:1995wm,Aldazabal:1995yw,Berkooz:1996iz,A,Choi:2002fn,Honecker:2006qz}. This provides not only a new top-down way to study vacua of orbifold compactification in the presence of heterotic 5-branes, opening more possibility for realistic vacua for the Standard Model, but also to understand global structure of moduli space. 

Instantons on non-compact orbifold $\R^4/\Z_N$, which provides a local geometry at the orbifold fixed points of $T^4/\Z_N$, provides a flat (constant) background connection at infinity \cite{Intriligator:1997kq,Berkooz:1996iz,A}. This information is parameterized by so-called shift vector $V$, which enters as an input in the string worldsheet CFT.  It is an outstanding observation in Ref. \cite{A} that the spectrum in the presence of heterotic 5-branes can be still be calculated using the same CFT just by modifying the zero point energy, proportional to the number of heterotic 5-branes. However there have been bottom-up approaches, in which quantitative factors are empirically determined and anomaly cancellation was used for consistency condition. In this paper we seek top-down approach, enabling justification on the calculation formulae and systematic classification of vacua. 

In the heterotic orbifold theories, the worldsheet CFT is exactly solvable \cite{DHVW} so that we can track the change of spectrum. This solvability includes the spectrum in twisted sectors in which the fields are localized at the orbifold fixed points, which is singular and cannot be accessed by field theory. In stringy calculation shift vector $V$ determines the twisted field spectrum thus we may hope that the above decomposition is possible and we can explain the non-perturbative vacua in terms of perturbative ones. Roughly speaking, using transition from perturbative orbifold vacua to non-perturvative ones, we can extract the information about small instantons from shift vectors by decomposing it, $V = V_1 + V_2$ into still broken part $V_1$ and instanton part $V_2$. The non-perturbative model is described by the new shift vector $V_1$.

An interesting consequence is that many vacua, which are regarded as disconnected described by different shift vector or instanton data, are connected by chains of phase transitions. It enables us to move from one vacuum to another. It also follows that orbifold fixed points also behave like branes and we also have transition among these three objects. Whether given string vacua are connected is an important problem. On one hand, to what extent the Standard Model is unique or preferred. We can seek a mechanism for dynamical selection of vacua.

\subsection{Global consistency condition}

To obtain realistic vacua, we break the symmetry, especially gauge symmetry, of heterotic string by associating it with the symmetry of internal space. In orbifold compactification, this information will be parameterized by a shift vector and we obtain spectrum by its projection that we review shortly.

The low energy limit of heterotic string is described by supergravity and the supersymmetry preserving solution is the semi-stable instantons \cite{Candelas:1985en}. The gauge symmetry is broken by this instanton background: If the structure group of the instantons is $G$, then that unbroken group is the commutant $H$ of $G$ in the mother group of heterotic string, $E_8\times E_8$ or $\text{\it Spin\,}(32)/\Z_2$ (which we usually call $SO(32)$ for convenience). What the shift vector describes should be related to this instanton background. It is also subject to the following global consistency condition.

The global consistency condition for the smooth K3 manifold is given by the Bianchi identity for the heterotic rank two antisymmetric $B$-field 
\begin{equation} \label{Bianchi}
  dH= d^2B +  \frac{1}{8\pi^2} \tr R^2  - \frac{1}{8\pi^2} \Tr F^2 = 0 . 
\end{equation}
Here $\Tr$ is the trace over adjoint representation of the gauge group, normalized by the corresponding dual Coxeter number. Also $R$ and $F$ are curvature two forms for the Lorentz and gauge connections, respectively, and the power is short for the wedge product, e.g. $F^n \equiv \wedge^n F$. 

If we take a background such that $d^2  \overline B = 0$,
the last term in (\ref{Bianchi}) under the instanton background $\Tr \overline F^2/(8\pi^2)$ is, integrated over a compact manifold, integrally quantized instanton number $k$. The orbifold geometry fixes the second term to be $\tr \overline R^2/(8\pi^2)$ and is integrated to give 24, the Euler number of the K3, so that 
\be \label{pertconsistency}
  24 - k =0.
\ee
For $E_8\times E_8$ heterotic string, the contribution on the gauge field can be decomposed into that of each $E_8$ and give $k=k_1+k_2$, with obvious notations.

We can regard a K3 as a blown-up $T^4/\Z_N$ orbifold, whose details we will study in Section \ref{sec:smallinst}.
In the orbifold limit, there appear localized fields at the orbifold fixed points. We will also see that the instanton number in (\ref{Bianchi}) can be decomposed into the bulk contribution $ \overline F_\U$ and fixed point contribution $ \overline F_\T$. The total sum should be preserved so that
\begin{equation} \label{Bianchitwisted}
  \frac{1}{8\pi^2} \tr  \overline R^2  -  \frac{1}{8\pi^2} \Tr  \overline F_\U^2 - \frac{1}{8\pi^2} \Tr  \overline F_\T^2= 0 . 
\end{equation}
Integrating over the entire $T^4/\Z_N$, the relation (\ref{Bianchitwisted}) is translated into the relations among numbers
\be \label{perturbativebianchi}
 24-  k_\U - k_\T =0.
\ee
The second and the third terms in (\ref{Bianchitwisted}) can be multiples of a fractional number, since the spectrum is (equally if there is no Wilson lines) distributed to a number of fixed points. For instance, in $T^4/\Z_2$ orbifold, Euler number $\frac{3}{2}$ is equally distributed on the sixteen fixed points \cite{Eguchi:1980jx}. In fact the untwisted sector contribution is {\em localized} at the fixed points as well, in the orbifold limit. 

Now we introduce a heterotic 5-brane. It provides magnetic source to $B$-field. Supposing  that they fill in the six dimensional uncompact space and look point-like in the internal space, so they satisfy the equation of motion
\be \label{heterotic5source}
 dH = \sum_a \delta^{(4)}(x-x_a),
\ee
which is Bianchi identity (\ref{Bianchi}) modified to \cite{Sagnotti:1992qw,Seiberg:1996vs,Choi:2017luj}
\begin{equation} \label{heterotic 5bianchi}
   \frac{1}{8\pi^2} \tr \overline R^2 - \sum_{a=1}^n \delta^{(4)}(x-x_a) - \frac{1}{8\pi^2} \Tr  \overline F^2= 0 ,
\end{equation} 
where $a=1,\dots,n$ labels different heterotic 5-branes. Integrating this over the entire orbifold, we obtain a consistency condition \cite{Seiberg:1996vs,Choi:2017luj}
\be \label{totalfive}
 24 - n - k = 0. 
\ee
We will explain the small instanton phenomena shortly, which can explain transfer between two numbers $n$ and $k$. In this paper we make extensive use of this to study the global structure of orbifold vacua.

Also we will consider more general combination of (\ref{perturbativebianchi}) and (\ref{totalfive}). For this it would be suggestive to rewrite the Bianchi identity as
\be \label{magnetic}
 d^2B =  \sum_{a=1}^n \delta^{(4)}(x-x_a) + \frac{1}{8\pi^2} \Tr F_\U^2 + \frac{1}{8\pi^2} \Tr F_\T^2  - \frac{1}{8\pi^2} \tr R^2 ,
\ee
where we now kept the fluctuation of $B$-field. There are some notable things.
The first three terms on the right-hand side have the same sign. It is discussed above that the first two terms can be transferred, preserving the total sum, but modifying individual $k_\U$ and $n$. Once we have the third term, the natural question is whether there can be more general transition among twisted fields, instantons and heterotic 5-branes as well. Another observation is that not only the first term, as in (\ref{heterotic5source}), provides the magnetic source to the $B$-field, but the rest of the terms provide also the same kind of sources. If they become delta functions by `shrinking' in the sense that we will review next, they can be regarded exactly the same object as the heterotic 5-branes.

\subsection{Small instantons}

An instanton, which we need for the reason discussed above, has continuous size. It may shrink to zero size becoming `small instanton.' Although the resulting profile becomes singular, the broken group, identical to structure group $G$, is restored \cite{Witten:1995gx,MV}. Small instantons are codimension four and undergo phase transition into  heterotic 5-branes \cite{Witten:1994tz}. The behavior is slightly different in two heterotic string theories.

In the $SO(32)$ heterotic string, there emerges extra gauge group $Sp(1) \simeq SU(2)$ for each small instanton. We have also a hypermultiplet transforming as a bifundamental under $H \times SU(2)$; the multiplicity is $\frac12$ due to half representation. 
This can be easily understood in type I dual, in which the spacetime of the $SO(32)$ theory and heterotic 5-branes are respectively mapped to D9/O9 and D5 branes. Each heterotic 5-brane gives rise to $Sp(1)$ gauge group due to the projection from O9 and when $n$ heterotic 5-branes become all coincident it is enhanced to $Sp(n)$. At the intersection between D9/O9 and D5, we have the bifundamental representation under $SO(32) \times Sp(n)$ group or its subgroup, from an open string connecting between them. Six dimensional gauge anomaly is nontrivial and it provides an important clue in understanding its non-perturbative structure.

In the $E_8\times E_8$ string, there is no additional gauge group from the 5-branes, but there is another interesting new effect. We can go to strong coupling limit to have M-theory with a new dimension \cite{HW,MV}. Two $E_8$'s are separately localized at the end of the interval, which are interpreted as another 9-branes. There can be a tensor branch in which such heterotic 5-branes are pulled out into the bulk and become M5-branes \cite{Ganor:1996mu}.

In what follows, we will apply this phenomenon to strings on toroidal orbifold $T^4/\Z_N$, formed by a discrete rotational element of $\Z_N \subset SU(2)$ holonomy on the torus. They can be regarded as a singular limit of the K3 manifold. In the smooth limit, instantons are spread over the internal manifold, as vector bundles. 
When we have orbifold limit, embedded instantons are localized at the fixed points. Their data are usually described by a shift vector $V$, is also explained by instantons \cite{Honecker:2006qz,Nibbelink:2007pn,Nibbelink:2007rd,Intriligator:1997kq}, in which references the transition between smooth and singular manifold is described. In this limit the structure group is the Cartan subalgebra corresponding to non-vanishing components of $V$. 

In the previous works, emphases have been made on the effect of heterotic 5-branes or instantons separately and the phase transition was not applied. In this paper we apply the phase transition idea to understand the connection between them. When there is a phase transition making small instantons into heterotic 5-branes we can track the change in the shift vector. This provides a top-down approach to understand the effect of non-perturbative effect.

\section{Orbifold vacua}

First we review perturbative orbifold vacua. The stringy spectrum is obtained by CFT with orbifold projection guided by modular invariance. However we can gain physical understanding on what is happening in terms of instantons. 

\subsection{Perturbative spectrum}

We compactify heterotic string on toroidal orbifold $T^4/\Z_N$ to yield six non-compact dimensions. We are particularly interested in prime cases $N=2,3$.
With the complexified coordinates $z^1=x^6+e^{2\pi i /3} x^7, z^2 = x^8+e^{2\pi i /3} x^9$ in the torus, the order $N$ twist acts as
\be \label{twist}
 (z^1,z^2) \to ( e^{2\pi i \phi_1} z^1,  e^{2\pi i \phi_2} z^2), \quad \phi = \left(\frac{1}{N},-\frac{1}{N}\right).
\ee
This preserves half of the supersymmetry. 

The current algebra is described by the weight vector $P$.
We associate this rotation with a shift of the weight vector $P \to P+V$, where $V$ is the order $N$ shift vector \cite{DHVW}. Both $P$ and  $NV$ belong to sixteen dimensional even and self-dual lattice $\Gamma_{16}$ or $\Gamma_{8} \times \Gamma_8$ depending on the gauge theory $SO(32)$ or $E_8 \times E_8$ \cite{Choi:2006qh}. Then we have unbroken gauge group and spectrum which are invariant under the orbifold projection.

The untwisted sector spectrum is obtained by
\begin{align}
\text{gauge bosons: } & P \cdot V \equiv 0 \mod 1, \label{Ugauge} \\
\text{untwisted matter: } & P \cdot V \equiv \frac{\ell}{N} \mod 1, \quad \ell=1,\dots \left \lfloor \frac{N}{2} \right \rfloor. \label{Umatter}
\end{align}
With appropriate combinations from the right-movers
we can make orbifold invariant state. In particular, the right mover has the phase $\phi_1 = \pm \frac{1}{N}$ from the shift, thus the states with $P \cdot V \equiv \pm \frac{1}{N}$ modulo an integer can survive. This will be important property of six dimensional toroidal orbifold models.

\subsection{Perturbative modular invariance}

We require modular invariance of string partition function to divergence free, worldsheet reparametrization invariant theory. It is generated by so-called $S$ and $T$ generators of $SL(2,\Z)$ transformation of modular parameter \cite{DHVW}. Here we consider the invariance condition in the absence of heterotic 5-branes first. 

Modular invariance under $S$ requires the existence of twisted strings, which are closed string up to orbifold action. Later we also see that twisted strings share many properties as open strings, whose ending hypersurface is interpreted as branes. We have $\lfloor N/2 \rfloor$ twisted sectors. In the $j$-th twisted sector the mass shell condition for the string left mover is
\be \label{TwMass}
 \tilde L_0 =0 : \alpha' m_L^2 = \frac{(P+jV)^2}{2}+ \tilde \N^{(j)} + E_0^{(j)}
\ee
with the string tension $\alpha'$ and the oscillator number $\tilde \N^{(j)}$, which is shifted, in the $j$th twisted sector. This comes from the constraint of Virasoro generator $\tilde L_0=0$. It also includes the untwisted sector as $j=0$. The zero-point energy is given as
\be \label{zeroptE}
 E_0^{(j)}=-1+\frac12\sum_{a=1}^2  \phi_a^{(j)} \left(1- \phi_a^{(j)} \right).
\ee
Here, by $\phi_a^{(j)}$ we subtract an appropriate integer from $j \phi_a$ to let it lie in the interval $[0,1)$. Also with the right movers, satisfying
\be \label{TwMassR}
L_0=0: \alpha' m_R^2 = \frac{(s+j\phi)^2}{2}+  \N^{(j)} + E_0^{(j)}+\frac12,
\ee
with a spacetime $SO(8)$ weight vector $s$, we have localized fields at the fixed points. Here again $L_0$ is the Virasoro generator. With the right-mover, the whole state is subject to generalized GSO projection, 
\be \label{generalizedGSO}
 e^{2\pi i \big(\tilde \N - \N+(P +V) \cdot V - (s+\phi)\cdot \phi - \frac12(V^2 - \phi^2)\big)}
\ee
by requiring this phase to be one \cite{Ibanez:1987pj,Katsuki:1989bf,Choi:2006qh}.

Invariance under $T$ comes from the level matching condition \cite{DHVW}
\be \label{levelmatching}
 \tilde L_0 = L_0.
\ee
Consider the twisted sector. The condition (\ref{levelmatching}) needs the following necessary condition
\be
 \frac{(P+V)^2}{2} + \tilde {\rm N} -  \frac{(s+\phi)^2}{2} - {\rm N} - \frac12 = 0 \mod 1.
\ee
In general there is no solution for the weight vector $P,s$ for a given $V,\phi$, because most of the terms proportional to $\frac{1}{N}$ cannot cancel the terms $\frac12(V^2 - \phi^2)$ proportional to $\frac{1}{2N^2}$. It is necessary to require the former to be also proportional. This is so-called the modular invariance condition 
\begin{equation} \label{modinv}
 \frac{V^2}{2} - \frac{\phi^2}{2} \equiv 0 \mod \frac{1}{N}.
\end{equation}
And the resulting vacua are sufficiently anomaly free \cite{SW}. The GSO phase is trivial when the massless condition is satisfied for modular invariant perturbative models.

\subsection{Classification of perturbative vacua}

In the orbifold embedding, the broken gauge group is parameterized by a shift vector.
First consider $SO(32)$ heterotic string. The shift vector of a type\footnote{In our notation, the superscript in the shift vector denote the repeated entries.}
\be \label{vectorialZNshift}
 V = \frac{1}{N} \bigg(0^{n_0}~1^{n_1}~\dots ~(N-1)^{n_{N-1}} \bigg)
\ee
with
\be \label{sixteen}
 \sum_{j=0}^{N-1} n_j = 16
\ee
is called vectorial. This vector yields unbroken gauge group
\begin{align}  
 \Z_2&: SO(2n_0) \times SO(2n_1) \label{unbrokenZ2} \\
 \Z_3&: SO(2n_0) \times SU(n_1 + n_2) \times U(1), \label{unbrokenodd}
\end{align}
according to (\ref{Ugauge}). Note that the untwisted and twisted spectrum is the same if we use an equivalent shift vector under Weyl reflection accompanied by lattice translations. The gauge symmetry (\ref{unbrokenodd}) can be partially explained by successive applications: if $n_0 >0$, $\frac13 (0^{n_0}~1^{n_1}~2^{n_2}) \sim \frac13 (0^{n_0}~1^{n_1}~2^{n_2-2}~-2~-2) \sim  \frac13 (0^{n_0}~1^{n_1+2}~2^{n_2-2})$. 
In the $\Z_2$ case there is a gauge symmetry enhancement due to self-conjugate, giving (\ref{unbrokenZ2}). 

There is another set of shift vectors, where $NV$ is a spinorial representation of $\Gamma_{16}$. Up to Weyl reflection and lattice translation, it can be generally expressed as
\be 
 V = \frac{1}{N} \left (\frac{1}{2}^{m_1}~ \frac32^{m_2} ~\dots ~\frac{2N-1}{2}^{m_N} \right),
\ee
with
\be
 \sum_{j=1}^{N} m_j = 16.
\ee
In the special case of $\Z_3$ orbifold it can be always brought to vectorial shift by Weyl reflections accompanied by a lattice shift by $(\frac12^{16})$. There is no independent spinorial shift vector. In the $\Z_2$ orbifold however the spinorial shift vector cannot be reduced; it always gives the unbroken gauge group
$$
 \Z_2: U(16).
$$

For the $E_8\times E_8$ case, we form sixteen component vector by direct sum of two $\Gamma_8$ shift vector either vectorial or spinorial. In the $\Z_2$ and $\Z_3$ cases, the same argument as above tells us that the only vectorial shift vector combinations are sufficient
\be \label{vectorialZ3e8e8}
 V =  \frac1N(0^{n_0}~1^{n_1} ~ \dots ~(N-1)^{n_{N-1}})(0^{n_0'}~1^{n_1'}~\dots~(N-1)^{n_{N-1}^\prime})
\ee
with
\be \label{eighteight}
  \sum_{i=1}^{N-1} n_i =  \sum_{i=1}^{N-1} n_i' = 8.
\ee
It is because up to Weyl reflections, we have three inequivalent groups in the $\Z_2$ case
\begin{align}
 \frac12(0^8) &: E_8, \nonumber \\
 \frac12(1^2~0^6) &: E_7 \times SU(2), \label{Z2subgroups} \\
 \frac12(1^4~0^4) &: SO(16) , \nonumber  
\end{align}
and five in the $\Z_3$ case
\begin{align}
 \frac13(0^8) &: E_8, \nonumber \\
 \frac13(1^2~0^6) &: E_7 \times U(1), \nonumber \\
 \frac13(2~0^7) &: SO(14)\times U(1), \label{Z3subgroups} \\
 \frac13(2~1^2~0^5) &: E_6 \times SU(3), \nonumber \\
 \frac13(2~1^4~0^3) &: SU(9).\nonumber 
\end{align}

Considering modular invariance condition the possibility reduces further. For $SO(32)$ string with $N=3$, the condition becomes 
\be \label{modinvZ3so32}
 n_1 + 4 n_ 2 \equiv 2 \mod 6.
\ee
Without further symmetry breaking, e.g. by Wilson line, we have only five inequivalent modular invariant shift vectors for SO(32) \cite{Choi:2006qh}, as listed in Table \ref{t:Z3}  for which we have $n=0$. For $E_8\times E_8$ string we have a similar condition,
\be
 n_1 +4 n_2 + n_1' + 4n_2' \equiv 2 \mod 6,
\ee
yielding also five combinations listed in Table \ref{t:Z3ofE8} with $n=0$. 
The shift vector for inequivalent, perturbative vacua are classified \cite{Choi:2004wn,Honecker:2006qz,Dixon,Choi:2003pqa}. Thus the forms of the shift vectors (\ref{unbrokenodd}) and (\ref{vectorialZ3e8e8}) are sufficiently general in the $\Z_3$ case. We can do similar classification, yielding the spectrum in Table \ref{t:t4z2so32} and Table \ref{t:t4z2e8e8}.

\subsection{ Instantons on $\R^4/\Z_N$ orbifold} \label{sec:smallinst}

The effect described by orbifold projection is captured by instanton background in the field theory limit  \cite{Intriligator:1997kq,A}. The gauge group from heterotic string is broken by this instanton background.

Each fixed point of $T^4/\Z_N$ orbifold is locally described by $\R^4/\Z_N$ and is regarded as singular limit of the $A_{N-1}$ ALE space \cite{Eguchi:1980jx,Nibbelink:2007rd}, which we discuss first. In the field theory limit, the following constant background cannot be gauged away due to orbifolding
\be
 A =  E \sum_{I=1}^{16} V_I H_I.
\ee
Here $E$ is the exceptional divisor from the blown-up singularity, satisfying self-intersection relation $E \cdot E= -N$.
The shift vector $V$ (\ref{vectorialZNshift}) has eigenvalues of the generator of the $SO(32)$ group or $SO(16) \times SO(16)$ maximal torus subgroup of $E_8\times E_8$ carried by heterotic string \cite{Intriligator:1997kq}. They are generated by Cartan subalgebra elements $H_I$'s.

In the orbifold limit, the unbroken group is enhanced to the group (\ref{unbrokenodd}) of rank sixteen \cite{Honecker:2006qz,Nibbelink:2007rd}, but the total instanton number should be preserved.
Inside $V$ in (\ref{vectorialZNshift}) we have $n_i$ eigenvalue for each $i/N$ \cite{Intriligator:1997kq,Berkooz:1996iz}\footnote{We use a different convention $w_m = 2 n_m$ than that in the reference.}
\begin{align} 
 k_{\Z_N} &= \frac{1}{8 \pi^2} \int  \Tr F^2 \\
   &  = \frac{N}{2} \sum_{i=0}^{N-1} \frac{i}{N}\left(1-\frac{i}{N}\right) n_i + K_N , \label{ZNALEinstno} \\
   & = \frac{N}{2} \sum_{I=1}^{16}  V_I(1-V_I) + K_N, \label{instnoshiftvec} 
\end{align}
where and in what follows the integration is done over the $A_{N-1}$ ALE space. Here the first term in (\ref{ZNALEinstno}) is the usual second Chern class from the flat connection from the infinity. Roughly speaking, in $F^2$ we have $V_I$ contribution on one $F$ and $1-V_I$ contribution on the other $F$ from the structure of the generator $H_I$ \cite{Berkooz:1996iz,Nibbelink:2007rd}. We may interpret that $K_N$ in (\ref{ZNALEinstno}) is the index of the localized fields at the tip of the ALE space. It can be an arbitrary number in the field theory, but the CFT of the global $T^4/\Z_N$ orbifold below completely determines it.

Now we wish to collect all the above local contributions to that the compact $T^4/\Z_N$ orbifold. To this end, we should know the multiplicity of fixed points of possible orders of $N$.
It is contained in the weights $d_N$ are the number of $\Z_{N}$ invariant fixed points, determined by global structure of $T^4/\Z_N$ by relating Euler numbers of fixed points
\be
 24 = \chi(T^4/\Z_N) =  d_N \chi(\R^4/\Z_N), N=2,3, \quad \chi(\R^4/\Z_\ell) = \frac{\ell^2-1}{\ell}.
\ee
Therefore we have the total instanton numbers 
\begin{align}
 k_\U = \frac{1}{8 \pi^2} \int \Tr F_\U^2 &=  d_N \frac{N}{2} \sum_{I=1}^{16}  V_I^{(j)}(1-V_I^{(j)}), \\
 k_\T = \frac{1}{8 \pi^2} \int \Tr F_\T^2 &=  d_N K_{N}, \label{twistedinstno}
\end{align}
which are valid for $N=2,3.$
Here again $V_I^{(j)}$ is $jV_I$ subtracted by an appropriate lattice vector to lie it in the interval $[0,1]$.
In the presence of Wilson lines, the degeneracy from $d_N$ disappears and we have independent contribution from local shift vectors at each fixed point.

For a $\Z_N$ shift vector of $SO(32)$ in (\ref{vectorialZNshift}), we have contributions at each fixed point
\be \label{Z3twinst} \begin{split}
 \Z_2: k_{\Z_2} &= K_2 + 2 \cdot \frac12 \cdot \frac12 \cdot \frac12 \cdot n_1,\\
 \Z_3: k_{\Z_3} &= K_3 + 3 \cdot \frac12 \cdot \frac13\cdot \frac23 \cdot (n_1+n_2) . 
\end{split}
\ee
In the absence of Wilson lines, all the fixed points are equal, thus in the bulk we have
\be \label{Z3instno} \begin{split}
 \Z_2: k &= 16 K_2 + 4 n_1, \\
 \Z_3: k &= 9 K_3 + 3 n_1 + 3 n_2.
\end{split}
\ee
Note that the instanton number is a gauge invariant quantity, as it should be. Either components $\frac13$ or $\frac23$ both describes $SU(n)$.

For $E_8\times E_8$ string, the shift vector plays the same role: each component parameterizes the rotation in the Cartan subalgebra direction. Thus with the shift vector (\ref{vectorialZ3e8e8}), we may use the same formula (\ref{instnoshiftvec}) for each $E_8$
\be \label{Z3twinstE8} \begin{split}
 \Z_2: k_{\Z_2} &= K_2 +  \frac14 n_1,\\
 \Z_3: k_{\Z_3} &= K_3 + \frac13 (n_1+n_2) ,
 \end{split}
\ee
where we count only the states charged under the first $E_8$. In the untwisted sector, usually a field is charged under one $E_8$. In the twisted sector, if a field is charged under the both, the representation becomes multiplicity. We have essentially the same counting for the second $E_8$. To sum up, we have
\be \label{Z3instnoE8E8} \begin{split}
 k & = k_1 + k_2,\\
 \Z_2: k_1 & = 16 K_2 + 4 n_1, \ k_2= 16 K_2' + 4n_1'\\
 \Z_3: k_1 & = 9 K_{3}+  3 n_1 + 3 n_2 ,\  k_2  = 9 K_{3}'+ 3 n'_1 + 3 n'_2.
\end{split}
\ee

This establishes the relationship between the modular invariance condition (\ref{modinv}) and Bianchi identity (\ref{Bianchi}).  We note that, although the condition (\ref{modinv}) has similarity with the Bianchi identity (\ref{Bianchi}), their contents are different. The shift vector and twist vector contain information of {\em breaking} of gauge symmetry and Lorentz symmetry, respectively, whereas the Bianchi identity contains the information about the spectrum. They are naturally related because the shift/twist vectors completely determine the spectrum.

\subsection{Chirality of localized fields} 

In the smooth compactification, all the 24 small instantons reside in the bulk geometry of internal manifold. However in the toroidal orbifold, we have twisted fields localized at the fixed points and they carry instanton number $k_\T$ in (\ref{twistedinstno}) as well.

There exists no index theorem that automatically gives the spectrum in the twisted sector, yielding $k_\T$, because field theory cannot take into account the stringy nature of the twisted strings. We should calculate the twisted fields from CFT formulae (\ref{TwMass}) and (\ref{TwMassR}). 

Interestingly, for the SO(32) string, anomaly consideration of $SO(2n_0)$ gauge group is sufficient to fix the instanton number $k_\T$ for the twisted strings.

In the twisted sector, the contribution from twisted fields to the instanton number $k_\T$ depends on the type $\bf R$ of localized fields. Essentially it is determined by the twisted sector spectrum. Let $\bf R$ be representations of $SO(2n_0)$. Then the anomaly is related to the coefficient $x_{\bf R}$ in
$$
 \tr_{\bf R} F^4 = x_{\bf R} \tr_{\rm v} F^4 + y_{\bf R} (\tr_{\rm v}F^2)^2,
$$ 
where $\tr_{\rm v}$ is over the vectorial representation of $SO(2n_0)$.
To sum up, in the perturbative model we have
\be
 k_\T =\sum_{{\bf R} \text{ for twisted fields}} x_{\bf R}.
\ee

This cancels six dimensional $SO(2n_0)$ gauge anomaly, with the chiral matter $\bf (n_1+n_{N-1},2n_0)$ under $SU(n_1+n_{N-1},2n_0)$ (or $\bf (2n_1,2n_0)$ of $SO(2n_1,2n_0)$ for $\Z_2$)
\be \label{kT}
  (2n_0 - 8) -n_1 - n_{N-1} -k_\T =0,
\ee
as long as  the total instanton numbers should satisfy the condition (\ref{perturbativebianchi})
$$  k = k_\U + k_\T = 24. $$
It is because we can show the following using (\ref{sixteen}) and (\ref{Z3instno})
\be
\begin{split}
 k_\T &= 24 - k_\U \\
    &= -8 + 32 - 2 \sum_{i=1}^{N-1} n_i - n_1 - n_{N-1} \\
    & = - 8 +2 n_0 - n_1 -  n_{N-1}.
 \end{split}
 \ee
This proves the relation (\ref{kT}).  The reason is that the total instanton number is related to the unbroken gauge group. Here the only unbroken part is $SO(2n_0)$.

In understanding the consistent vacua, the $SO(2n_0)$ anomaly plays an important role.
In six dimension, the chirality of gaugino in the vector multiplet is always opposite to the fermions in the hypermultiplets, so that the matter contents are constrained by anomalies.
Six dimensional gauge anomalies can be cancelled, up to Green--Schwarz mechanism, if the anomaly polynomial has vanishing $\tr F^4$ term \cite{Intriligator:1997kq}.
From the contributions of gaugino
\begin{align}
 \Tr F^4_{SO(2n)} &= (2n-8) \tr_{\rm v} F^4_{SO(2n)} + 6 (\tr F^2_{SO(2n)})^2 ,\label{SOanomaly}\\
 \Tr F^4_{SU(n)} &=2n \tr F^4_{SU(n)} + 6 (\tr F^2_{SU(n)})^2 ,
 \end{align}
we see thaat we need $(2n-8)$ vectors and $2n$ fundamentals, respectively, to cancel  $SO(n)$ and $SU(n)$ anomalies. Other representations may contribute, as we have summarized the decomposition in the Appendix. 

The instanton number is different from other anomaly coefficients, for example of $SU(n)$-type subgroups, or exceptional subgroups of $E_8$, or even $SO(n_1)$ of the $\Z_2$ orbifold.
However there is instanton number contribution for $k_\T$ from matter states. 

$E_8\times E_8$ heterotic string vacua have a number of distinct features. There is no relation between anomaly and instanton numbers. Technically the zero entries of the shift vector do not yield $SO(n_0)$ but  different groups listed in (\ref{Z2subgroups}) and (\ref{Z3subgroups}). Here the instanton contribution comes from vectorial component of the representation under the branching in the maximal $SO(n_0)$ group. For instance, $\bf 27$ of $E_6$ contributes 1 and $\bf 56$ of $E_7$ contributes 2 to the instanton number $k_\T$. We have no gauge symmetry enhancement from heterotic 5-branes, and any $E_n$ subgroups are anomaly free, thus the anomaly freeness is less restrictive.

\section{Non-perturbative vacua and their transitions}

Now we introduce heterotic 5-branes. As alluded in the introduction, there is phase transition between small instantons and 5-branes, so that different vacua can exchange instantons. It turns out that many seemingly unconnected vacua are actually connected. Since the effect of small instantons is parameterized by the shift vector, we can trace information from the difference of the shift vectors of vacua  and we can understand how they are connected by instanton exchange. From this we can prove many relations among the zero point energy, the number of 5-branes and difference of the shift vectors. 

\subsection{Small instantons into heterotic 5-branes}

The information about instanton embedding is encoded in the shift vector $V$. Some of instantons can be emitted into bulk to become heterotic 5-branes, recovering larger unbroken group. The resulting unbroken group should be also described by another shift vector that we shall call $V_1$. Decomposing 
\be \label{shiftdecomposition}
 V = V_1+V_2,
\ee
the emitted instanton components are described by $V_2$, which describes also the recovered part.

We first study the mass shell condition. Consider the twisted sector of a {\em perturbative,} modular invariant theory parameterized by $V$. Expanding the mass shell condition (\ref{TwMass}), we have
\be \label{mshellEcorr}
\begin{split}
\frac12 \alpha' m_L^2  & = \frac{(P+V_1)^2}2+ \tilde \N +P\cdot V_2 + V_1 \cdot V_2  + \frac12 V_2^2+ E_0 \\
 & = \frac{(P+V_1)^2}{2}+ \tilde \N + E_0 + \Delta E_0.
 \end{split}
\ee
We may regard this formula as the mass shell condition for the twisted sector for a daughter vacuum with $V_1$, with instantons emitted. Then the extra piece from the expanding $V$ can be regarded as modification in the zero-point energy
\be \label{zeroptEcorr}
 \Delta E_0 \equiv V_1 \cdot V_2 + \frac{1}{2} V_2^2.
\ee
A part of instantons described by the shift vector $V_2$ `condensates' in the CFT description. This justifies the shift of the zero point energy in Ref. \cite{A}.
We assumed $P \cdot V_2=0$ because we want $\Delta E_0$ is a constant in the new, daughter vacuum, not depending on $P$.

The spectrum can still be calculated using the same CFT in the presence of 5-branes, using the same formula as that of the perturbative case \cite{A}
$$
 m_L^2 = 0.
$$
This is possible because we have inherited the vacuum from a perturbative vacuum.
Since the inclusion of heterotic 5-branes does not affect the internal geometry of orbifold, we have no change in the spacetime part, including that of the right mover. 

We can also have modified, generalized GSO projector (\ref{generalizedGSO}) in the presence of heterotic 5-branes. Since the terms involving $V$ is modified as 
\be \begin{split} 
 (P+V) \cdot V - \frac12 V^2 & = (P+V_1)^2 - \frac12 V_1^2 + \frac12 V_2^2 + V_1 \cdot V_2 \\
 	& =  (P+V_1)^2 - \frac12 V_1^2 + \Delta E_0,
\end{split}
\ee
where again $P \cdot V_2 = 0$ is assumed. Therefore, we have modified, generalized GSO projector for a non-perturbative shift vector $V$ taking into account heterotic 5-branes
\be 
 e^{2\pi i \big(\tilde \N - \N+(P +V) \cdot V - (s+\phi)\cdot \phi - \frac12(V^2 - \phi^2) + \Delta E_0 \big)}.
\ee
The extra phase is simply expressed by $\Delta E_0$. The 5-branes can only affect the CFT on the fixed point only quantitatively. Note that the combination (\ref{zeroptEcorr}) is invariant under Weyl reflection. Indeed it is reflection that obviously preserves the inner product.

\subsection{Modified modular invariance condition} \label{sec:NPmodinv}

The change of mass shell condition affects the modular invariance.
As in the perturbative case, we may consider level matching condition with the modified mass shell condition (\ref{mshellEcorr}), now with a correction to the zero-point energy, that might take into account non-perturbative effects
\be
 \frac{(P+V)^2}{2}+ \tilde \N + \Delta E_0  -  \frac{(s+\phi)^2}{2} - {\rm N} + \frac12 = 0 \mod 1.
\ee
Since each term is proportional to $1/N$, we expect their cancellation gives an integer sum, expect the $V^2$ and $\phi^2$ terms. Thus we require the modified modular invariance condition
\be \label{modinvNP}
 \frac{V^2}{2} + \Delta E_0 -  \frac{ \phi^2}{2} \equiv 0  \mod \frac{1}{N}.
\ee
In the perturbative case we have $\Delta E_0=0$ giving the previous condition (\ref{modinv}), but this fails in this case. Instead we require the form (\ref{modinvNP}). 

A shift vector $V$ that fails to satisfy {\em perturbative} modular invariance condition (\ref{modinv}) can nevertheless describe a consistent non-perturbative vacuum \cite{A}.  In the $SO(32)$ heterotic string theory, a failure of anomaly cancellation can be remedied by extra chiral fields coming from the heterotic 5-branes. In the $E_8\times E_8$ heterotic string, most vacua are anomaly free even without perturbative modular invariance.

We obtained the modular invariance condition (\ref{modinvNP}) as a necessary condition. In the perturbative case that condition (\ref{modinv}) also turns out to be sufficient condition. Unfortunately, this is not the case for the non-perturbative case. Even though we had shift vectors  satisfying the condition we sometimes have anomalous vacua. We will come to a further selection rule and counterexample in Section \ref{sec:selrule}.

The modular invariance condition has preferred basis, so not every Weyl equivalent shift vector satisfies the same modular invariance condition. The modular invariance condition (\ref{modinvZ3so32}) is not symmetric under the exchange $n_1$ and $n_2$, although the unbroken group (\ref{unbrokenodd}) is. It is rather a condition about the absolute size of the shift vector. The twisted sector mass shell condition (\ref{TwMass}) depends on it.

\subsection{Example} \label{s:hettrans}

Through this section, we consider an exemplar transition
\begin{itemize}
\item from the perturbative vacua of $G=U(8)\times SO(16)$ with the shift vector $V = \frac13(1^8~0^8)$ in Table \ref{t:Z3}.
\item to the non-perturbative vacua $G_1=U(2) \times SO(28)$ with the shift vectot  $V_1 = \frac13(1^2~0^{14})$ in Table \ref{t:Z3}.
\end{itemize}

The shift vector $V$ of $G$ is decomposed into $V_1$ of $G_1$ and an extra $V_2$ as
\begin{equation} \label{exampledecomposition}
 V = V_1 + V_2 = \frac13(1^2~0^{14}) + \frac13(0^2~1^6~0^{8}).
\end{equation}
Upon phase transition the small instanton components described by $V_2$ is going to become heterotic 5-branes. Therefore we are left with the remaining component $V_1$, describing a non-perturbative vacua with the gauge group $G_1$.

The unbroken gauge group is $U(2) \times SO(28)$, with untwisted matter ${\bf (2,28)}+3{\bf (1,1)}$, due to instantons embedded in the structure group $\Z_3 \subset U(1)\times SU(2)$. It is described by the shift vector $V_1$, which is eventually related to the instanton number $k_\U=3n_1=6$. 

In the twisted sector, on top of the usual zero-point energy 
\be 
 E_0 = -1 + 2 \cdot \frac12 \cdot \frac13 \cdot \left(1-\frac13 \right) = - \frac{7}{9},
\ee
there is modification as in (\ref{zeroptEcorr}), 
\be \label{ex1}
 \Delta E_0 = 0 + \frac{1}{2} \left( \frac{1}{3} (1^6 ~ 0^{10})\right)^2 = \frac{18}{54}. 
\ee
Plugging these into the mass shell condition (\ref{mshellEcorr}), we find the spectrum shown in Table \ref{t:Z3}. There is no charged representation. Since there is no $SO(28)$ vector or spinor, thus no instanton contribution comes from the twisted sector $k_\T=0$. 

As a result, the 18 instantons described by $V_2$ is completely converted to as many $n=18$ heterotic 5-branes by phase transition, so that the total number is conserved to be 24 as in (\ref{totalfive}). The biggest gauge group $Sp(18)$ is generated if all the heterotic 5-branes become coincident. Besides the vector multiplets of both groups $SO(28)$ and $Sp(18)$, we have hypermultiplets in $({\bf 2,1;36})$ and $\frac12({\bf 1,28;36}).$
With the total 20 vectors, we can check that gauge anomalies of $SO(28)$ cancel from (\ref{SOanomaly}). 

In the above example, $V_1 \cdot V_2=0$ happened, which can be relaxed in the following. There can be a transition between the perturbative $U(5)\times SO(22)$ model to non-perturbative $U(4)\times SO(24)$ model via
\be \label{decomp1}
\begin{split}
 V & = \frac13 (2 ~1^4 ~0^{11}) = V_1 + V_2 \\
 V_1 & = \frac13 (1~1~1~1~0~0^{11}) \\
 V_2  &= \frac13 (1~0~0~0~1~0^{11}).
 \end{split}
\ee
This gives the modification of zero-point energy $\Delta E_0 = V_1 \cdot V_2 + \frac12 V_2^2 = \frac{2}{9}$. This also lead us to anomaly-free vacuum. Alternatively, we can take another combination
\be \label{decomp2}
\begin{split}
 V_1' & = \frac13 (0~1^4~0^{11})\\
 V_2' & = \frac13 (2~0^{15}).
\end{split}
\ee
Remarkably the zero-point energy is corrected in the same amount $\Delta E_0 = V_1' \cdot V_2' + \frac12 V_2^{\prime 2} = \frac{2}{9}$. The two models have the same spectrum. As long as $V_1$ and $V_1'$ are related by Weyl transformation, we should always have the same $\Delta E_0$.

\subsection{Change of spectrum and anomaly flow}

Let us analyze the spectrum change during the phase transition. First we see that the vector representations (in the hypermultiplets) in the untwisted sector of $G_1$ can be explained simply by branching from those of $G$. We may consider common subgroup 
$$H \equiv U(2)\times U(6) \times SO(16)$$
which is a subgroup of both $G$ and $G_1$. From the branching $G \to H$, we see 
\begin{align}
 ({\bf 8,16}) &\to {\color{olive} ({\bf2,1,16})} + ({\color{violet}\bf 1,6,16}), \\
 ({\bf 28,1}) &\to ({\color{violet} \bf 1,15,1}) + {\color{olive} (\bf 2,6,1)} + ({\bf 1,1,1}).
\end{align} 
From here we can see that $SO(16)$ vector can only come from the multiplet $ ({\bf 8,16}) $ that is present in the untwisted sector. We expect from the gauge symmetry enhancement $H \to G_1 = U(2) \times SO(28),$ we have recombination
\begin{align}
 \color{olive} ({\bf 2,28}) &\leftarrow  \color{olive} ({\bf2,6,1}) + ({\bf 2,\overline 6,1}) + ({\bf 2,1,16}), \\
 \color{violet} ({\bf 1,378}) &\leftarrow \color{violet} ({\bf 1,36,1}) + ({\bf 1,1,120}) + ({\bf 1,15,1}) + ({\bf 1,\overline {15},1}) \nonumber \\ & \quad \color{violet} + ({\bf 1,6,16}) + ({\bf 1,\overline 6,16}). 
\end{align} 

Following the recombination, the eight $\bf 16$ of $SO(16)$ underwent the recombination into two $\bf 28$ of $SO(28)$. Out of 24 small instantons, 18 of them come out into the bulk.
Therefore the chirality is simply branching and recombination of vector fields in the untwisted sector. In particular only heterotic 5-brane can give rise to the vector of $SO(28)$.

We can also keep track of anomaly flow during phase transition.
We have derived the instanton number (\ref{Z3instno}) from the field strength. Here, we can understand the multiplicative factors $3$ in front of $n_1$ and $n_2$ in another way by looking at the spectrum change.

The emitted instantons are described by a part of the shift vector $\Delta V = \frac{1}{3}( 0^p~ 1^q~ 0^r)$ where $p+q+r=16$ (In the above case, we have $p=2,q=6,r=8$). Since $U(p+q)$ became $U(p)$, 
the number of vectors reduces by $q$ and thus the number of necessary heterotic 5-branes for the anomaly cancellation of SO-type group increase by $2q$. Thus the decreased small instanton is $3q$, which should explain the coefficients 3 in (\ref{Z3instno}). Due to even root property of weight vector, $q$ should be multiple of 2 thus the number of extracted instantons should be always a multiple of 6.

The same argument is possible for the transition from $SO(2n+4) \times SO(2m) \to SO(2n) \times SO(2m+4)$ in the $\Z_2$ orbifold. The emitted instantons are described by $\Delta V = (\frac12~ \frac12~ 0^{14})$ up to permutation. The number of vector reduces by 4 but the necessary instantons increase by 4.

\subsection{Dual and inverse transitions}

The above phase transition always has a dual process. From the decomposition $V = V_1+V_2$, if we take a daughter model using the shift vector $V_2$, the emitted instanton data is contained in $V_2$, thus the zero-point energy is
\be \label{revE}
  \Delta_{\text{dual}} E_0 = V_2 \cdot V_1 + \frac{1}{2} V_1\cdot V_1.
\ee

For this dual transition, let us come to the example with (\ref{exampledecomposition}).
Exchanging the role of $V_1$ and $V_2$, we can view the transition as that of  $U(8)\times SO(16) \to U(6) \times SO(20) \equiv G_2$ vacua. Six small instantons now described by $V_1$ may undergo transition into the same number of heterotic 5-branes.
Accordingly, another recombination takes place
\begin{align}
 ({\bf 6,20}) &\leftarrow ({\bf 1,6,16}) +({\bf 2,6,1}) + ({\bf \overline 2, 6,1}) , \\
 ({\bf 15,1}) &\leftarrow ({\bf 1,15,1}) \\ ({\bf 1,190}) &\leftarrow ({\bf 4,1,1}) + ({\bf 1,1,120}) + ({\bf 2,1,16}) + ({\bf \overline 2,1,16}) + 2({\bf 1,1,1}) . 
\end{align}
Again, we verify that the modified zero-point energy (\ref{revE}) 
\be \label{ex2}
 \Delta E_0 = V_2 \cdot V_1 + \frac12 \left(V_1\right)^2 = \frac19 = \frac{6}{54}. 
\ee
This verifies also the transition of small instantons into $n=12$ heterotic 5-branes. It should be noted that, however, for one transition $V_1$ yielding the same vacua, there can be many different dual vacua, as in the examples (\ref{decomp1}) and (\ref{decomp2}).

Reverse process, from a daughter theory with $V_1$ above to the mother theory with $V$, is also possible. If we express
\be
 V_1 = V - V_2
\ee
then we can change the shift vectors of the mother and the daughter theories 
\be \begin{split}
 \Delta_{\text{rev}} E_0 &= - V \cdot V_2 + \frac{1}{2} V_2^2 \\
  &= - V_1 \cdot V_2 -\frac{1}{2} V_2^2 \\
  &= - \Delta E_0.
  \end{split}
\ee
Thus the process should be reversible.

\subsection{The combination: Twisted states to heterotic 5-branes} \label{s:tw2heterotic 5}

For perturbative models we always have the relation $k = k_\U + k_\T =24$ as in (\ref{perturbativebianchi}). If we take heterotic 5-branes into account, this condition is relaxed to $k+n =24$ as in (\ref{totalfive}). The phase transition takes place changing these number but preserving the sum $\Delta k + \Delta n =0$. 

Here we consider another transition between twisted field and 5-branes
\be \label{twisttransition}
 \Delta k_\T + \Delta n = 0, \quad k_\U + k_\T + n = 24.
\ee
Since we have no control over the twisted sector fields via field theory, we cannot prove the direct transition giving (\ref{twisttransition}). However, noting the $k_\U$ is completely determined by the shift vector, we can indirectly use the following chain of transitions; If a shift vector at hand $V$ can be embedded into another perturbative shift vector $V'$ as
\be \label{twistedtransition}
 V' = V + V_2,
\ee
then we can apply the transition between small instantons (encoded in $V_2$) and 5-branes, having a different twisted sector.

Such example can be found in Table \ref{t:Z3}.
The {\em non-perturbative} vacuum we have obtained from the transition in Section \ref{s:hettrans} has the shift vector 
$$ V = \frac13(1^2~0^{14}). $$
However there is also a {\em perturbative} vacuum with the same shift vector satisfying modular invariance condition (\ref{modinv}).
The entire untwisted sector is the same in both cases, because we use the identical conditions for mass and projection. The gauge group is $U(2)\times SO(28)$ and the instanton number $k_\U = 3 n_1 = 6$ are the same. 

The modification to zero-point energy affects the twisted field spectrum.
The $k_\T$ essentially counts the number vector multiplets $\bf 28$ or equivalent in the twisted sector. We have $SU(2)$ doublet in each of nine fixed points. Without Wilson line we do not distinguish among them. Therefore in the perturbative vacuum, we remove $2 \times 9= 18$ vectors from the twisted sector. This should give rise to the same number of vectors from heterotic 5-branes in the non-perturbative model, satisfying (\ref{twisttransition}).

\begin{figure}[h]
\begin{center}
$$
\begin{tikzcd}[row sep=5em]
 & \text{(small instantons)} \arrow[ddr,leftrightarrow]{} \\
 \\
\text{(twisted fields)} \arrow[uur,leftrightarrow]{} \arrow[rr,leftrightarrow]{} && \text{(heterotic 5-branes)}
\end{tikzcd}
$$
\end{center}
\caption{There are phase transitions among small instantons, heterotic 5-branes and twisted fields. \label{f:transitions}}
\end{figure}
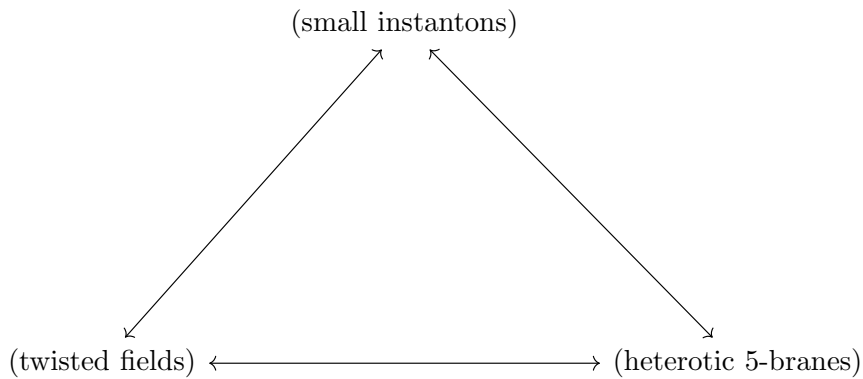

In all the above processes, the total number (\ref{totalfive}) is always preserved
\be
 k_\U +  k_\T + n = 24 .
\ee
This is the result of the combination of all the processes: We have transition among small instantons, twisted states at orbifold fixed points, heterotic 5-branes as in the Figure \ref{f:transitions}.

The transition between fixed states and 5-branes happens in general: if we have a transition between small instantons, changing $k_\U$, to heterotic 5-branes, there is also change in the twisted sector spectrum in general, changing $k_\T$. It is because the twisted spectrum is changed since the mass shell condition is changed as in (\ref{mshellEcorr}).

We can view these exchanges as dynamics of 5-branes. The dual description of small instantons are D5-branes on top of D9-branes. In fact, the fixed points can also be regarded as 5-branes. They are localized $(5+1)$-dimensional hypersurface on which the twisted strings are localized \cite{Forste:2004ie,Kobayashi:2004ud}, in a similar sense that open strings that are localized on a NS5 or D5-brane. Moreover, we have seen that Eq. (\ref{pertconsistency}) can be regarded as the magnetic equation for the $B$-field, in which the instantons and fixed points are all the magnetic sources.

\subsection{Selection rule} \label{sec:selrule}

We have seen phase transitions that are understood of exchange of 5-branes, in which the important constraint is the preserved number of 5-branes as in the condition (\ref{totalfive}). In addition, this phase transition in {\em orbifold theory} takes place in a special form, as in Eq. (\ref{mshellEcorr}), namely that should be parameterized by the shift vector. Therefore the transition is further constrained by a selection rule that we shall see here.

Let us calculate the modification in the zero-point energy of the $\Z_3$ vacua. They are described by shift vectors (\ref{vectorialZNshift}).
The instanton number is given in (\ref{Z3instno})
\be
 k = 9 K_3 + 3 n_1 + 3 n_2  = 24 - n.
\ee
There should be $n$ of heterotic 5-branes subject to global consistency condition (\ref{totalfive}).
We have supposed that this vacuum is inherited from a perturbative modular invariant theory with
\be
 k^\m = 9 K_3^\m + 3 n_1^\m + 3 n_2^\m = 24,
\ee
where the superscripts indicate that the corresponding quantities are those of perturbative theory.
The change of each contribution is described as follows,
\begin{align*}
 \Delta n_i = n_i - n_i^\m, \quad i=1,2, \quad  \Delta K_3 = K_3 - K_3^\m.
\end{align*}
Note that, the change of instanton number is not only due to the change of the components $V_1$ but also the chirality of twisted sector spectrum. When small instantons get transition into heterotic 5-branes, both are related. Subtracting them we have
\be \label{instnochange}
  9\Delta K_3+ 3 \Delta n_1 + 3 \Delta n_2 + n = 0. 
\ee

In the simplest case where the shift vector is simply decomposed $V=V_1 + V_2$ and there is no overlapping elements, then the change of zero-point energy is obtained from (\ref{zeroptEcorr})
\be \label{zeroptEchange}
 \Delta E_0 = \frac{1}{2} \left( - \Delta n_1 \left( \frac{1}{3} \right)^2 - \Delta n_2 \left( \frac{2}{3} \right)^2 \right).
\ee

We may relate the number $n$ of heterotic 5-branes in (\ref{instnochange}) and the change $ \Delta E_0 $ of the zero point energy (\ref{zeroptEchange}) as follows. In case $\Delta n_1 \ne 0$ we may eliminate it to have 
\be \label{Ederivation}
 \Delta E_0 = \frac{n - 9 \Delta K_3 - 9 \Delta n_2}{6 N^2},
\ee
with $N=3$ here.
This vacuum should be independent of transition, thus the modified zero point energy (\ref{Ederivation}) should only be dependent on the number of heterotic 5-branes.  It follows
\be \label{Z3selrule}
  \Delta n_2 +  \Delta K_3 = 0.
\ee
We may have this same non-perturbative vacua from {\em another perturbative} vacuum with $\Delta n_2 \ne 0$. However everything should be same if the transition is commutative. This means that we should always have 
\be
 n = - 9 \Delta K_3 - 3 \Delta n_2 - 3 \Delta n_1 =  - 3 \Delta n_1 - 12 \Delta n_2 
\ee
and
\be \label{zerpotEproptoheterotic 5}
 \Delta E_0 = \frac{n}{6 N^2}.
\ee
regardless of $\Delta n_1 =0$ or not. Thus, the relation (\ref{Z3selrule}) provides a selection rule. This is nontrivial because, for a given shift vector, we cannot expect what kind of twisted fields can be obtained. In particular, if there is no change in the chirality in the twisted sector $\Delta K_3=0$, then the instanton described by entries with $n_2 > 0$ cannot be changed.

This selection rule (\ref{Z3selrule}) states that some transition is not possible. For instance, the phase transition from the perturbative vacuum with $V=\frac13(1^{14}~0^2)$ to a non-perturbative with $V_2=\frac13(1^{12}~0^4)$ is not possible. There is no change in $\Delta n_2$ thus we need $\Delta K_3=0$ from the rule (\ref{Z3selrule}). A stringy calculation using $V_2$ shows that we would have a $U(12)\times SO(8)$ vacuum with the twisted sector fields $9{\bf (66,1)}+18{\bf (1,8_s)}+18{\bf (1,1)}$ contributing $k_\T=-9=9 \Delta K_3 \ne 0$. Note that we have no way to predict the change and rearrangement of the twisted fields in the field theory limit so the selection rule is necessary condition. The resulting spectrum is anomalous, too. The would-be chain involving a non-perturbative vacuum with $V=\frac13(1^{10}~0^6)$ is forbidden.
This also shows that not every model satisfying the modified modular invariance condition (\ref{modinvNP}) is not a sufficient condition in the non-perturbative vacua.
The main reason is the anomaly contribution of spinorial representations. If we obtained the vacuum from the transition, the spinor representation cannot be well-branched under the transition (technically it can be seen from the mass formula in the twisted sector (\ref{TwMass}). If we have only vector representations in the twisted sectors, it is likely that we have well-defined non-perturbative vacua.

This selection rule is based on some assumptions. First, assumed that all the fixed points are equivalent. This may be relaxed if we have Wilson lines. There can be a further complication due to the possibility of $V_1 \cdot V_2 \ne 0$. Also we assumed that there is no overlapping components between $V_1$ and $V_2$; Otherwise the modification of the zero point energy $\Delta E_0$ has different dependence on $\Delta n_i$'s. We will come back to this in the example below. Since this selection rule relates the number of instantons and the shift of zero point energy, it holds valid for $E_8 \times E_8$ vacua.

\section{Connected vacua}

\begin{table}[t]
\footnotesize
\begin{center}
\begin{tabular}{ccc}
\hline \hline
Shift vector $V$ & Untwisted &$k_{\rm U}$  \\
 \multirow{2}{*}{Group} & Twisted & $k_{\rm T}$ \\
 & heterotic 5 localized & $n$ \\
  \hline 
$\frac13(0^{16})$  & $2{\bf (1;1)}$ & 0  \\
 \multirow{2}{*}{$SO(32)\times Sp(24)$} & $18 {\bf (1;1)}$ & 0 \\
 & $\frac12{\bf (32;48)}$ & 24 \\
 \hline
$\frac13(2~0^{15})$ & ${\bf( 30;1)}+2{\bf (1;1)}$ &  3 \\
  \multirow{2}{*}{$U(1)\times SO(30)\times Sp(12)$} & $9{\bf( 30;1)}+18{\bf( 1;1)}$ & 9 \\
 & $\frac12{\bf (30;24)}$  & 12 \\ 
 \hline
$\frac13(1^2~0^{14})$ & ${\bf (2,28;1)}+2{\bf (1,1;1)}$ &  6 \\
  \multirow{2}{*}{$U(2)\times SO(28)\times Sp(18)$} & $9 {\bf (1,1;1)}+18 {\bf (1,1;1)}$ & 0 \\
 & $\frac12{\bf (1,28;36)}+{\bf (2,1;36)}$ & 18 \\ 
\hline
$\frac13(1^2~0^{14})$ & ${\bf (2,28)}+3{\bf(1,1)}$ & 6 \\
 \multirow{2}{*}{$U(2)\times SO(28)$} &  $9{\bf (2,28)}+63{\bf(1,1)}$ & 18\\
& & 0 \\
\hline
$\frac13(2~1^2~0^{13})$& ${\bf (3,26;1)}+{\bf(3,1;1)}+2{\bf (1,1;1)}$  & 9 \\
  \multirow{2}{*}{$U(3)\times SO(26)\times Sp(6)$} & $9{\bf (1,26;1)}+18{\bf (3,1;1)}$  & 9 \\
&  $\frac12{\bf (1,26;12)}+{\bf (3,1;12)}$ & 6 \\
\hline
$\frac13(1^4~0^{12})$ & ${\bf (4,24;1)}+{\bf (6,1;1)}+2{\bf (1,1;1)}$  &12 \\
  \multirow{2}{*}{$U(4)\times SO(24)\times Sp(12)$} & $9{\bf (6,1;1)}+18{\bf (1,1;1)}$& 0 \\
&  $\frac12{\bf (1,24;24)}+{\bf (6,1;24)}$ & 12 \\
\hline
$\frac13(2~1^4~0^{11})$ & ${\bf (5,22)}+{\bf(10,1)}+2{\bf(1,1)}$   & 15  \\ 
 \multirow{2}{*}{$U(5)\times SO(22)$} & $9{\bf (1,22)}+9{\bf(10,1)}+18{\bf (5,1)}$ & 9 \\
 & & 0\\
 \hline
$\frac13(1^6~0^{10})$ &  ${\bf (6,20;1)}+{\bf (15,1;1)}+2{\bf (1,1;1)}$ & 18   \\
  \multirow{2}{*}{$U(6)\times SO(20)\times Sp(6)$}& $9{\bf (15,1;1)}+18{\bf (1,1;1)}$ &  0 \\
  & $\frac12{\bf (1,20;12)}+{\bf (6,1;12)}$ & 6\\
\hline
$\frac13(1^8~0^{8})$ & ${\bf (8,16)}+{\bf(28,1)}+2{\bf(1,1)}$ & 24\\
 \multirow{2}{*}{$U(8)\times SO(16)$} & $9{\bf (28,1)}+18{\bf (1,1)}$   & 0 \\ 
 & & 0 \\ 
 \hline
%{\color{red} $\frac13(1^{10}~0^6)$} & ${\bf (10,12)}+{\bf (45,1)}$ & 30  \\
% $U(10)\times SO(12) \times SO(12)$  & $9{\bf (1,1)}$ & 0 \\
% &$ { (\bf 10,1;12)}+\frac12{(\bf 1,12;12)}$ & $-6$ \\
% \hline
$\frac13(2~1^{10}~0^{5})$&  ${\bf (11,10)}+{\bf (55,1)}+2{\bf(1,1)}$  & 33 \\
 \multirow{2}{*}{$U(11)\times SO(10)$} & $9{\bf (11,1)}+9{\bf (1,16)}$ & $-9$ \\
& & 0\\ 
\hline
% $\frac13(1^{12}~0^4)$& ${\bf (12,8)}+{\bf (66,1)}+2{\bf (1,1)}$  & 36 \\
%$\color{red} U(12)\times SO(8) \times SO(6)$ & $18{\bf (1,8_s)}+18{\bf (1,1)}$ & $-9$ \\
% & & $-3$ \\
% \hline
 $ \frac13(1^{14}~0^2)$ & ${\bf (14,4)}+{\bf(91,1)}+2{\bf(1,1)}$  & $42$ \\
\multirow{2}{*}{$U(14)\times SO(4)$} & $9{\bf (1,1)}+9{\bf (14,2)}+18{\bf (1,2)}$ & $-18$ \\
& & 0 \\
\hline
%{\color{blue} $\frac13(\frac{3}{2}^2~\frac{1}{2}^{14})$} &  ${\bf (14,4)}+{\bf(91,1)}+2{\bf(1,1)}$  &   $24$ \\
%\multirow{2}{*}{$U(14)\times SO(4)$} & $9{\bf (1,1)}+9{\bf (14,2)}+18{\bf (1,2)}$ & $0$ \\
%& & 0 \\
%\hline
%$\frac13(-\frac32\frac12^{15})$ & ${\bf (15,1;1)}+\bf (\overline{15},1;1)+{\bf (105,1;1)}$ &18 \\
%  \multirow{2}{*}{$U(15) \times SO(2) \times Sp(6)$} &  $9 {\bf (15,1;1)} + 27{\bf (1,1;1)}$& 0\\
%&  $\frac12{\bf (1,2;12)}+{\bf (15,1;12)}$ &6 \\
%\hline 
%$\frac13(\frac12^{16})$ & ${\bf (120;1)}$  &12 \\
 % \multirow{2}{*}{$U(16)\times Sp(12) $} & $18 {\bf (1;1)}$& 0 \\
% &  ${\bf (16;24)}$ &12 \\
%\hline
\end{tabular}
\end{center}
\caption{Perturbative ($n=0$) and non-perturbative ($n \ne 0$) vacua of $SO(32)$ string on $T^4/\Z_3$ orbifold. The parameters $k_\U,k_\T,n$ are respectively instanton numbers in the untwisted and twisted sectors, and the number of heterotic 5-branes. $k_{\rm U} = 3 m$ with the rank of the $U(m)$ factor group, and $k_{\rm U} + k_{\rm T} +n = 24$. \label{t:Z3}}
\end{table}

\begin{table}[t]
\footnotesize
\begin{center}
\begin{tabular}{ccc}
\hline \hline
Shift vector $V$ & Untwisted &$(k_1,k_2)_\U$  \\
 \multirow{2}{*}{Group} & Twisted & $(k_1,k_2)_\T$ \\
 & heterotic 5 localized & $n$ \\
  \hline 
$\frac13(0^8)(0^8)$  &  $2 {\bf (1,1)}$ &$(0,0)$  \\
 \multirow{2}{*}{$E_8 \times E_8 $} & $18 {\bf (1,1)}$ & $(0,0)$  \\
 && 24\\
 \hline
$\frac13(1^2~0^6)(0^8)$ & ${\bf( 56,1)}+3{\bf (1,1)}$ & $ (6,0)$ \\
  \multirow{2}{*}{$E_7 \times U(1)\times E_8$} & $9{\bf(56,1)}+18{\bf( 1,1)}$ & $(18,0)$ \\
 & & 0 \\ 
  \hline
$\frac13(1^2~0^6)(0^8)$ & ${\bf( 56,1)}+3{\bf (1,1)}$ & $ (6,0)$ \\
  \multirow{2}{*}{$E_7 \times U(1)\times E_8$} & $27{\bf (1,1)}$ & $(0,0)$ \\
 & & 18 \\
 \hline
 $\frac13(2~0^7)(0^8)$ & ${\bf (14,1)}+{\bf (64,1)}+2{\bf (1,1)}$ & $ (3,0)$ \\
  \multirow{2}{*}{$SO(14)\times U(1)\times E_8$} & $9{\bf(14,1)}+18{\bf( 1,1)}$ & $(9,0)$ \\
 & & 12 \\ 
  \hline
  $\frac13(0^8)(2~1^2~0^5)$ & ${\bf (1,27,3)} + 3{\bf (1,1)}$ & $ (0,9)$ \\
  \multirow{2}{*}{$E_8\times E_6 \times SU(3)$} & $9{\bf(1,27,1)}+18{\bf(1,1,3)}$ & $(0,9)$ \\
 & & 6 \\ 
 \hline
 $\frac13(1^2~0^6)(2~0^7)$ & ${\bf( 56,1)}+{\bf (1,14)} + {\bf (1,64)}+3{\bf (1,1)}$ & $ (6,3)$ \\
  \multirow{2}{*}{$E_7 \times U(1)\times SO(14) \times U(1)$} & $9{\bf(1,14)}+9{\bf (1,1)}+18{\bf(1,1)}$ & $(0,9)$ \\
 & & 6 \\ 
 \hline
$\frac13(1^2~0^6)(2~1^2~0^5)$ & ${\bf( 56,1,1)}+{\bf (1,27,3)} + 3{\bf (1,1)}$ & $ (6,9)$ \\
  \multirow{2}{*}{$E_7 \times U(1)\times E_6 \times SU(3)$} & $9{\bf(1,27,1)}+18{\bf(1,1,3)}$ & $(0,9)$ \\
 & & 0 \\ 
 \hline
$\frac13(1^2~0^6)(1^2~0^6)$ & ${\bf( 56,1)}+{\bf(1,56)}+4{\bf (1,1)}$ & $ (6,6)$ \\
  \multirow{2}{*}{$E_7 \times U(1)\times E_7 \times U(1)$} & $18{\bf( 1,1)}+18{\bf( 1,1)}$ & $(0,0)$ \\
 & & 12 \\ 
  \hline
$\frac13(2~1^4~0^3)(0^8)$ & ${\bf( 84,1)}+2{\bf (1,1)}$ & $ (15,0)$ \\
  \multirow{2}{*}{$SU(9)\times E_8$} & $9{\bf(36,1)}+18{\bf(9,1)}$ & $(9,0)$ \\
 & & 0 \\ 
 \hline
  $\frac13(2~0^7)(2~0^7)$ & ${\bf (14,1)}+{\bf (64,1)}+{\bf (1,14)}+{\bf (1,64)}+2{\bf (1,1)}$ & $ (3,3)$ \\
  \multirow{2}{*}{$SO(14)\times U(1)\times SO(14)\times U(1)$} & $9{\bf(14,1)}+9{\bf(1,14)}+18{\bf( 1,1)}$ & $(9,9)$ \\
 & & 0 \\
 \hline
 $\frac13(2~1^2~0^5)(2~1^2~0^5)$ & ${\bf(27,3,1,1)}+{\bf (1,1,27,3)}+2{\bf (1,1,1,1)}$ & $ (9,9)$ \\
  \multirow{2}{*}{$E_6 \times SU(3) \times E_6 \times SU(3)$} & $9{\bf(1,3,1,3)}$ & $(0,0)$ \\
 & & 6 \\ 
 \hline
  $\frac13(2~1^4~0^3)(2~1^2~0^5)$ & ${\bf( 84,1,1)}+{\bf (1,27,3)}+2{\bf (1,1,1)}$ & $ (15,9)$ \\
  \multirow{2}{*}{$SU(9)\times E_6 \times SU(3)$} & $9{\bf(9,1,3)}$ & $(0,0)$ \\
 & & 0 \\ 
 \hline
  \end{tabular}
\end{center}
\caption{Perturbative ($n=0$) and non-perturbative ($n \ne 0$) vacua of $E_8\times E_8$ heterotic string on $T^4/\Z_3$ orbifold.  \label{t:Z3ofE8}}
\end{table}

\begin{table}[t]
\footnotesize
\begin{center}
\begin{tabular}{ccc}
\hline \hline
Shift vector $V$ & Untwisted &$k_{\rm U}$  \\
 \multirow{2}{*}{Group} & Twisted & $k_{\rm T}$ \\
 & heterotic 5 localized & $n$ \\
  \hline 
$\frac12(0^{16})$  & $2{\bf (1;1)}$ & 0  \\
 \multirow{2}{*}{$SO(32)\times Sp(24)$} & $18 {\bf (1;1)}$ & 0 \\
 & $\frac12{\bf (32;48)}$ & 24 \\
 \hline
$\frac12(1^2~0^{14})$  &  ${\bf (4,28)}+2{\bf(1,1)}$  & 8  \\
 \multirow{2}{*}{$SO(4)\times SO(28)$ } &  $8{\bf (2_c,28)}+16{\bf(2_s,1)}$ & 16 \\
 & & 0 \\
 \hline
 $\frac12(1^2~0^{14})$  &  ${\bf (4,28)}+2{\bf(1,1)}$  & 8  \\
 \multirow{2}{*}{$SO(4)\times SO(28)$ } &  $8{\bf (2_c,1)}$ & 0 \\
 & & 16 \\
 \hline
 $\frac12(1^4~0^{12})$  & ${\bf (8,24;1)}+2{\bf(1,1;1)}$ & 16  \\
 \multirow{2}{*}{$SO(8)\times SO(24)\times Sp(8)$} & $8{\bf (8_s,1;1)}$ & 0 \\
 & $\frac12{\bf (8,1;16)} + \frac12{\bf (1,24;16)}$ & 8 \\ 
 \hline
 $\frac12(1^6~0^{10})$  &${\bf (12,20)}+4{\bf(1,1)}$ & 24  \\
 \multirow{2}{*}{$SO(12)\times SO(20)$} & $8{\bf (32,1)}$  & 0 \\
 & & 0 \\
 \hline
 $\frac12(\frac12^{16})$  & $2{\bf (120;1)}+2{\bf(1;1)}$ &  16  \\
 \multirow{2}{*}{ $U(16)\times Sp(8)$} & $16{\bf (1;1)}$ & 0 \\
 & $\frac12{\bf (32;48)}$ & 8\\
  \hline
  $\frac12(\frac12^{15}~\text{-}\frac32)$ &$2{\bf (120)}+4{\bf(1)}$& 24  \\
 \multirow{2}{*}{$U(16)$} & $16{\bf (16)}$ & 0 \\
 & & 0 \\
 \hline
\end{tabular}
\end{center}
\caption{Perturbative $(n=0)$ and non-perturbative vacua of $SO(32)$ heterotic string on $T^4/\Z_2$ orbifold. $k_\U,k_\T,n$ are small instanton number, the number of heterotic 5-branes and the number of heterotic 5-branes. These are all the inequivalent vacua satisfying the modular invariance condition up to Weyl reflections. \label{t:t4z2so32}}
\end{table}

\begin{table}[t]
\footnotesize
\begin{center}
\begin{tabular}{ccc}
\hline \hline
Shift vector $V$ & Untwisted &$(k_1,k_2)_\U$  \\
 \multirow{2}{*}{Group} & Twisted & $(k_1,k_2)_\T$ \\
 & heterotic 5 localized & $n$ \\
  \hline 
$\frac12(0^{8})(0^8)$  & $2{\bf (1;1)}$ & (0,0)  \\
 \multirow{2}{*}{$E_8 \times E_8$} & $8 {\bf (1;1)}$ &$(0,0)$ \\
 &  & 24 \\
 \hline
$\frac12(1^2~0^6)(0^8)$  &  ${\bf (56,2)}+4{\bf(1,1)}$  & $(8,0)$  \\
 \multirow{2}{*}{$E_7 \times SU(2) \times E_8$ } &  $8{\bf (56,1)}+16{\bf(1,1)}$ & $(16,0)$ \\
 & & 0 \\
 \hline
 $\frac12(1^2~0^6)(0^8)$  &  ${\bf (56,2)}+4{\bf(1,1)}$  & $(8,0)$  \\
 \multirow{2}{*}{$E_7 \times SU(2) \times E_8$ } &  $8{\bf (56,1)}+16{\bf(1,1)}$ & $(0,0)$ \\
 & & 16 \\
 \hline
 $\frac12(1^2~0^6)(1^2~0^6)$  &  ${\bf (56,2,1,1)}+{\bf (1,1,56,2)}+2{\bf(1,1,1,1)}$  & $(8,8)$  \\
 \multirow{2}{*}{$E_7 \times SU(2) \times E_7 \times SU(2)$ } &  $8{\bf (1,2,1,2)}$ & $(0,0)$ \\
 & & 8 \\
 \hline
 $\frac12(0^8)(1^4~0^4)$  &  ${\bf (1,128)} + 2{\bf(1,1)}$  & $(0,16)$  \\
 \multirow{2}{*}{$E_8 \times SO(16)$ } &  $8{\bf (1,16)}$ & $(0,0)$ \\
 &  & 8 \\
 \hline

 $\frac12(1^2~0^6)(1^4~0^4)$  &  ${\bf (56,2,1)}+{\bf (1,1,128)} + 2{\bf(1,1,1)}$  & $(8,16)$  \\
 \multirow{2}{*}{$E_7 \times SU(2) \times SO(16)$ } &  $8{\bf (1,2,16)}+16{\bf(1,1,1)}$ & $(0,0)$ \\
 & & 0 \\
 \hline
 \end{tabular}
\end{center}
\caption{Perturbative $(n=0)$ and non-perturbative vacua of $E_8 \times E_8$ heterotic string on $T^4/\Z_2$ orbifold. \label{t:t4z2e8e8} }
\end{table}

\begin{figure}[t]
\begin{tikzcd}[row sep=large,cells={nodes={draw,minimum height=0.6cm}}]
 & \parbox{4cm}{\centering$\frac{1}{3}(0^{16})$ \\ $SO(32) \times Sp(24)$ } \ar[d,leftrightarrow]  & 
\\
 &  \parbox{4cm}{\centering$\frac{1}{3}(1^2~0^{14})$  \\ $U(2)\times SO(28) \times Sp(18)$ }\ar[d,leftrightarrow] \ar[dr,leftrightarrow]  \\
 & \parbox{4cm}{\centering$\frac{1}{3}(1^4~0^{12})$ \\
  $U(4)\times SO(24)$ }\ar[dr,leftrightarrow] \ar[d,leftrightarrow] 
 & \parbox{4cm}{\centering$\frac{1}{3}(2~0^{15})$ \\ $SO(30)$ }  \ar[d,leftrightarrow] 
 \\
  & \parbox{4cm}{\centering$\frac{1}{3}(1^6~0^{10})$ \\ $U(6)\times SO(20)$ }  \ar[d,leftrightarrow] \ar[dr,leftrightarrow]
  & \parbox{4cm}{\centering$\frac{1}{3}(2~1^2~0^{13})$ \\ $U(3) \times SO(26)$ }  \ar[d,leftrightarrow] 
  \\
   \parbox{4cm}{\centering$\frac{1}{3}(1^2~0^{14})$ \\ $U(2)\times SO(28)$} \ar[uuur,dashed,leftrightarrow]  & \parbox{4cm}{\centering$\frac{1}{3}(1^8~0^{10})$ \\ $U(8)\times SO(16)$ }  
   & \parbox{4cm}{\centering$\frac{1}{3}(2~1^4~0^{11})$ \\ $U(5) \times SO(22)$ }
   \\
   & \parbox{4cm}{\centering$\frac{1}{3}(1^{14}~0^2)$ \\ $U(14)\times SO(4)$ }   \ar[r,dashed,leftrightarrow]
   & \parbox{4cm}{\centering$\frac{1}{3}(2~1^{10}~0^{3})$ \\ $U(11)\times SO(6)$ }
\end{tikzcd}
\caption{Connected vacua of $SO(32)$ heterotic string on $T^4/\Z_3$ orbifold. Among five perturbative models, bottommost one in each column, three generate chains of connected vacua. Only the nearest transitions with $\Delta E_0 = \pm \frac19$ is indicated by solid arrows. Transition for dashed arrows can be proven using the chain of such transitions.} \label{fig:transSO32}
\end{figure}
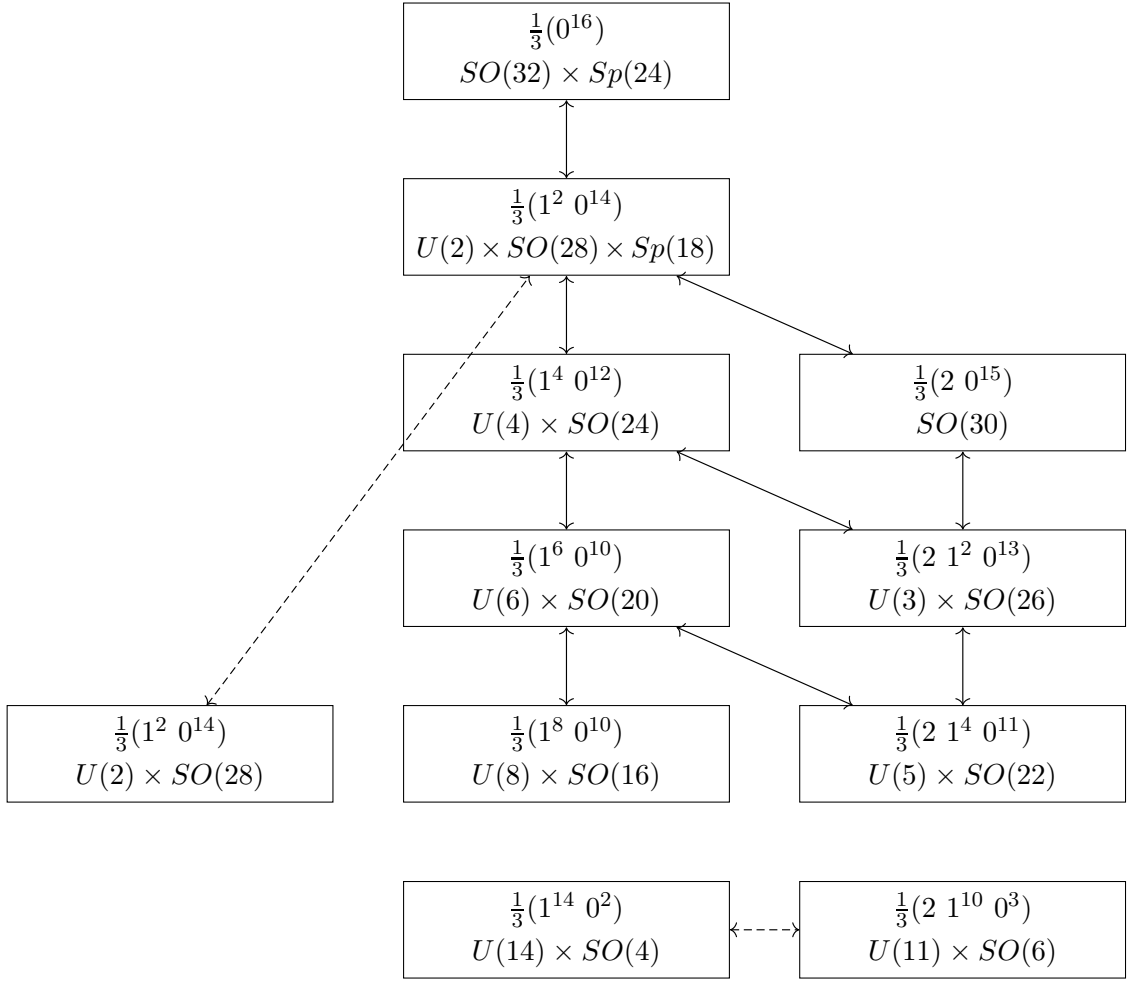

\begin{figure}[h]
\begin{tikzcd}[row sep=large,cells={nodes={draw,minimum height=0.6cm}}]
 & \parbox{4cm}{\centering$\frac{1}{3}(0^8)(0^8)$ \\ $E_8\times E_8$ } \ar[d,leftrightarrow]  & 
 \\
 &  \parbox{4cm}{\centering$\frac{1}{3}(1^2~0^6)(0^{8})$  \\ $E_7  \times U(1)\times E_7 \times U(1)$ }\ar[d,leftrightarrow] \ar[dr,leftrightarrow] 
 &   \parbox{4cm}{\centering$\frac{1}{3}(1^2~0^6)(0^{8})$ \\ $E_7 \times U(1)\times E_8$} \ar[l,Leftrightarrow] 
 \\
  &  \parbox{4cm}{\centering$\frac{1}{3}(1^2~0^6)(1^2~0^{6})$ \\ $SO(14)\times U(1) \times E_8$ }\ar[d,leftrightarrow] 
  & \parbox{4cm}{\centering$\frac{1}{3}(2~0^7)(0^8)$ \\ $SO(14)\times U(1) \times E_8$ } \ar[d,leftrightarrow]
\\
  & \parbox{4cm}{\centering$\frac{1}{3}(2~0^7)(1^2~0^6)$ \\ $SO(14)\times U(1) \times E_7 \times U(1)$ }  \ar[ur,leftrightarrow] 
  & \parbox{4cm}{\centering$\frac{1}{3}(2~1^2~0^5)(0^8)$ \\ $E_6 \times SU(3) \times E_8$ }  \ar[d,leftrightarrow]  \ar[dl,leftrightarrow]
\\   \parbox{4cm}{\centering$\frac{1}{3}(2~0^7)(2~0^7)$ \\ $SO(14)\times U(1) \times SO(14)\times  U(1)$ }  \ar[ur,leftrightarrow]
  & \parbox{4cm}{\centering$\frac{1}{3}(2~1^2~0^5)(1^2~0^6)$ \\ $SU(9) \times E_7 \times U(1)$ }    \ar[u,leftrightarrow]
  &  \parbox{4cm}{\centering$\frac{1}{3}(2~1^4~0^3)(0^8)$ \\ $SU(9)\times E_8$  } 
\\
  \parbox{4cm}{\centering$\frac{1}{3}(2~1^2~0^5)(2~1^2~0^5)$ \\ $E_6\times SU(3)\times E_6\times SU(3)$ }   \ar[d,leftrightarrow] & &
 \\
  \parbox{4cm}{\centering$\frac{1}{3}(2~1^4~0^3)(2~1^2~0^5)$ \\ $SU(9)\times E_6\times SU(3)$ }  & &
\end{tikzcd}
\caption{Connected vacua of $E_8 \times E_8$ heterotic string on $T^4/\Z_3$ orbifold. Among five perturbative models, bottommost one in each column, three generate chains of connected vacua.} \label{fig:transE8E8}
\end{figure}
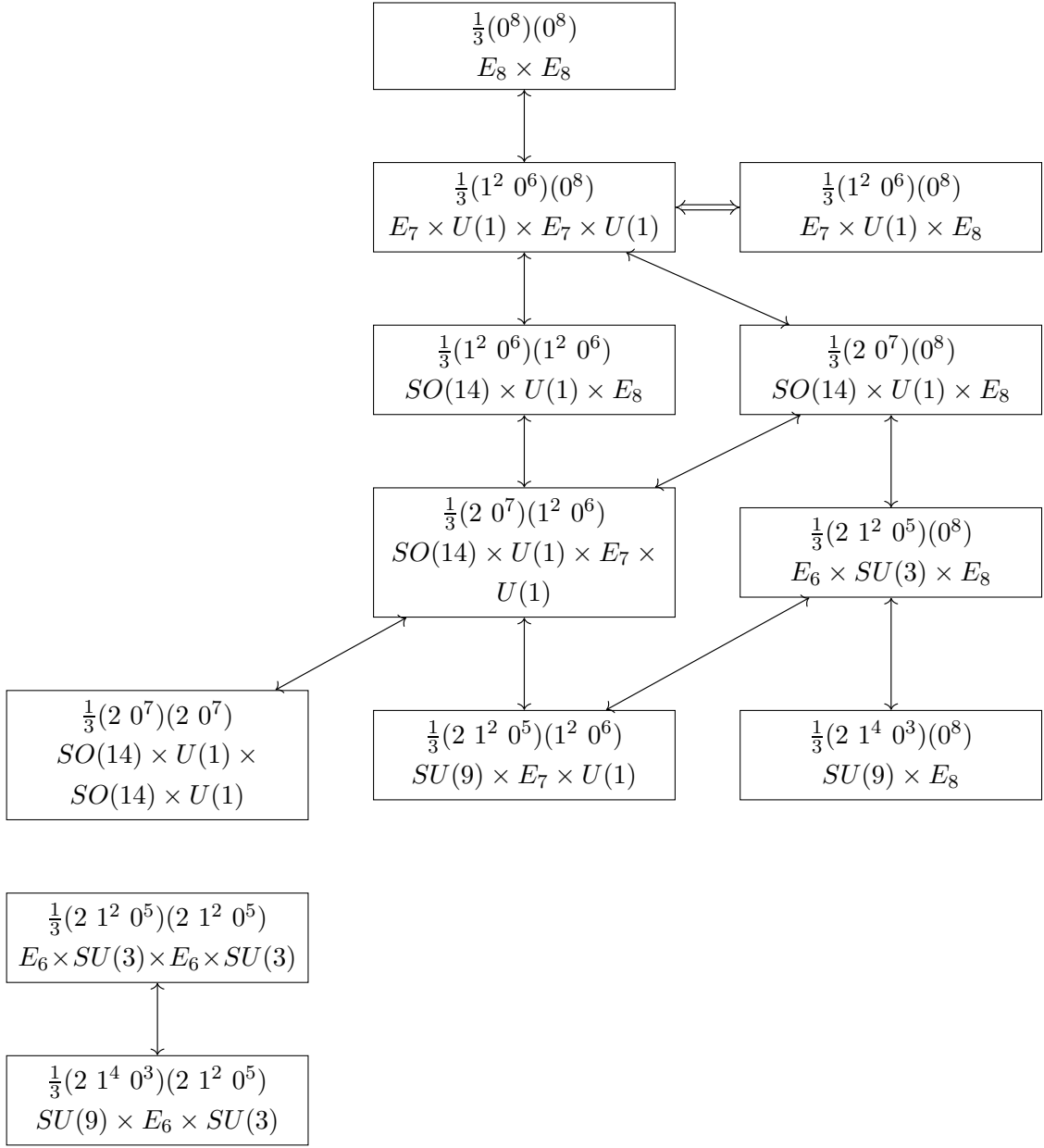

The above discussion forces us to conclude that most of the vacua are connected. Chains of transitions of small instantons, heterotic 5-branes and twisted fields takes one vacuum to another. All the $\Z_2$ and $\Z_3$ vacua of both heterotic string theories are shown in Tables \ref{t:t4z2so32} through Table \ref{t:Z3ofE8}.

\subsection{Connected vacua of $\Z_3$ orbifold}

The transition map for the $T^4/\Z_3$ orbifold is depicted in Figure \ref{fig:transSO32} for $SO(32)$ and in Figure \ref{fig:transE8E8} for $E_8\times E_8$. 
Each heterotic string on this orbifold has five perturbative, modular invariant vacua.  In $SO(32)$ vacua two of them are disconnected and admit no transitions because the instanton number of the untwisted sector is too large. In these models, we have large spinorial representations contributing the instanton number $k_\T$ in the twisted sector. After the transition, in the presence of heterotic 5-brane, we have large positive shift of the zero point energy, hence it is difficult to have still the spinorial representation, making transition impossible. 

Starting from one, we can go to another by emitting small instantons into heterotic 5-branes. 
We can travel every solid single arrow by transition parameterized by auxiliary shift vector 
\be \label{ejection}
 V_2=\frac{1}{3}(0^a~\pm1~\pm1~0^b), \quad a+b=14.
\ee
We may go directly to any of them by ejecting a multiple of six instantons at the same time. In particular in $SO(24)\times U(2)$ model there is a transition between twisted fields and heterotic 5-branes, whose direction is denoted by horizontal direction. In all of the above transitions, we have the relation between the shift vector and emitted instanton number
$$
 \Delta E_0 = V_1 \cdot V_2 + \frac12 V_2^2= \frac{\Delta n }{6 N^2}
$$
generalizing (\ref{zerpotEproptoheterotic 5}).
(For the transition from a mother theory with $V=V_1+V_2$ to a daughter theory with $V_1$ we have always $V_1 \cdot V_2=0$.)
The reverse transition is possible so the transition is two way. Note that there can be a direct transition between two vacua even if they are not directly connected by an arrow. For instance, we have seen that we can go directly from $\frac13(1^8~0^8)$ vacuum to  $\frac13(1^2~0^{14})$ vacuum.

This argument may hold for two vacua with the same instanton numbers. Consider for example $V=\frac13(1^4~0^{12})$ and $V_1=\frac13(2~0^{15})$ are horizontally connected by chains of transitions. We can describe their direct connection using $V_2 = \frac13(-1~1~ 1~ 1~ 0^{12})$, giving $\Delta E_0 = V_1 \cdot V_2 + \frac12 V_2^2 = 0$. If they are directly connected by virtual instanton exchanges, we may expect that the bottommost two vacua of SO(32), with $\frac13(1^{14}~0^2)$ and $\frac13(2~1^{10}~0^3)$  be connected by transition as well.

When all the 24 small instantons are emitted and become coincident we have the maximal gauge group $Sp(24)$. The biggest `bifundamental' representation is $(32,48)$ in the half multiplet. Note that in this case, the duality between $E_8\times E_8$ and $SO(32)$ works, because all the spectra have one-to-one correspondence. The brainy picture in type I side has been well-studied \cite{Berkooz:1996dw,Choi:2006hm,Choi:2006th}. 

For the $E_8 \times E_8$ models, we have two groups of connected vacua. In the $E_8$ case, the shift vector $V = \frac13(2~0^7)(0^8)$ is equivalent to $V' = \frac13(1^4~0^4)(0^8)$ by Weyl reflection. Starting from either vector we can obtain the descendent model with $V_1 = \frac13(1^2~0^6)(0^8)$ by respectively shift vectors $V_2=\frac13(1~-1~0^6)(0^8)$ and $V_2' = \frac13(0~0~1~1~0^4)(0^8)$. They all give the same zero point energy correction  $\Delta E_0 = V_1 \cdot V_2 + \frac12 V_2^2 = V'_1 \cdot V'_2 + \frac12 V_2^{\prime 2}$.

There are transitions between perturbative vacua, in which small instantons are exchanged to twisted fields. We have shown this using the chain of dualities.

Note that there is duality between two heterotic string theories. At the tip of the chains of each string theory, we have the same field contents. So the two heterotic string theories are dually related.

\subsection{Communication between two $E_8$'s}

In many cases, the selection rule (\ref{Z3selrule}) applies well to supplement the modular invariance condition. For example, transition from $V=\frac13(2~1^4~0^3)(2~1^2~0^5)$ to $V_1 = \frac13(2~1^4~0^3)(1^2~0^6)$ is not possible because while $\Delta n_2 =1$ we have $\Delta K_3 = 0$. However there are anomalous $E_8\times E_8$ vacua even they pass both conditions; The resulting spectrum is anomalous. They are listed in Table \ref{t:inconsistentE8}. 

For example, the model $V_1' = \frac13(2~1^4~0^3)(2~0^7)$ is anomalous although the transition from the vacuum with $V$ seems possible according to the selection rule from $\Delta K_3 = \Delta n_2=0$. It would be anomaly free if some twisted fields are branched from those in the vacuum with $V$. The desired representation does not arise in the twisted sector because the modification of the zero point energy is too large for this representation to satisfy the mass shell condition. 

Another example is a seemingly-consistent vacuum with 
\be \label{wedontknow}
 \frac13(2~1^2~0^5)(2~1^2~0^5) \text{ giving } E_6 \times SU(3)\times E_6 \times SU(3).
\ee 
It is not anomalous because the unbroken gauge groups are anomaly free. Note that this vacuum is an intermediate one along the chain of transition from $V=\frac13(2~1^4~0^3)(2~1^2~0^5)$ to the above inconsistent vacua, so this may be inconsistent as a string model although it is field theoretically anomaly free.

\begin{table}[t]
\footnotesize
\begin{center}
\begin{tabular}{ccc}
\hline \hline
Shift vector $V$ & Untwisted &$(k_1,k_2)_\U$  \\
 \multirow{2}{*}{Group} & Twisted & $(k_1,k_2)_\T$ \\
 & heterotic 5 localized & $n$ \\
  \hline
   $\frac13(2~1^2~0^5)(2~0^7)$ & ${\bf(27,3,1,1)}+{\bf (1,1,14,1)}+{\bf (1,1,64,1)}+2{\bf (1,1,1,1)}$ & $ (9,3)$ \\
  \multirow{2}{*}{$E_6 \times SU(3) \times SO(14)\times U(1)$} & $9{\bf(1,3,1,1)}$ & $(0,0)$ \\
 & & 12 \\ 
 \hline
  $\frac13(2~1^4~0^3)(2~0^7)$ & ${\bf( 84,1)}+{\bf (1,14)}+{\bf (1,64)}+ 2{\bf (1,1)}$ & $ (15,3)$ \\
  \multirow{2}{*}{$SU(9)\times SO(14) \times U(1)$} & $9{\bf(9,1)}$ & $(0,0)$ \\
 & & 6 \\ 
 \hline
    $\frac13(2~1^2~0^5)(2~1^2~0^5)$ & ${\bf(27,3,1,1)}+{\bf (1,1,27,3)}+2{\bf (1,1,1,1)}$ & $ (9,9)$ \\
  \multirow{2}{*}{$E_6 \times SU(3) \times E_6 \times SU(3)$} & $9{\bf(1,3,1,3)}$ & $(0,0)$ \\
 & & 6 \\ 
 \hline
  \end{tabular}
\end{center}
\caption{(Possibly) anomalous vacua of $E_8\times E_8$ heterotic string on $T^4/\Z_3$ orbifold, but passing the selection rule. \label{t:inconsistentE8} }
\end{table}

One interesting observation is that, except this one with  (\ref{wedontknow}), {\em all} the non-perturbative vacua in the presence of heterotic 5-branes contain {\em no charged field under two groups originating from different $E_8$'s.} The above anomalous model, which satisfy the modified modular invariance condition and pass the selection rules, requires charged field {\em under two groups coming from different} $E_8$'s. This provides a suggestive explanation on  the communication problem between two $E_8$'s, and deserves further study \cite{Gorbatov:2001pw}. By emitting small instantons in one $E_8$ and absorbing them in the other $E_8$, chiralities can be exchanged. Thus these two $E_8$ can be exchanged.

In the strong coupling limit we open up a new interval that separates these two $E_8$'s. The 5-brane is interpreted as M5-brane if we regard this theory as M-theory compactified on an interval and after phase transition this can be emitted in the bulk as explained in the introduction. Therefore M5-branes can be messengers for exchanging chirality between the two $E_8$'s. The technical reason making sense of this argument is that if we have non-zero number $n$ of heterotic 5-branes, this introduces modification of the zero point energy, which makes hard to satisfy nontrivial mass shell condition. In this situation, typically it is very difficult to have a charged state under two groups inherited from different $E_8$'s, which partly justify the above observation. This would let us to rule out the vacuum (\ref{wedontknow}). It would be also interesting for further study if this transition should always be involved in communicating two $E_8$'s in the strong coupling limit of $E_8 \times E_8$ heterotic string.

\section{Discussion}
 
We have studied vacua of heterotic string compactified on $T^4/\Z_N$ orbifold, with $N=2,3,$ in the presence of heterotic 5-branes. A new understanding comes from phase transition between small instantons and 5-branes. It does not only explain the  spectrum, but also suggest that most of vacua of toroidal orbifolds are connected. This hints at the evolution of stringy vacua in the landscape. 

In the presence of heterotic 5-branes, although the theory is non-perturbative, we have still exactly solvable CFT in the worldsheet. This is because the 5-branes can be obtained by instanton transition from a perturbative vacuum, which is indeed described by a solvable CFT. The effect of heterotic 5-brane modifies the zero-point energy \cite{A}. Rather than previous bottom-up approach, we can derive the modified zero-point energy using phase transition between small instantons and 5-branes. With this, we derive modified modular invariance condition. This gives us deeper understanding on the relation between the modular invariance and Bianchi identity.

The most striking feature is the connection between two perturbative vacua, in which some small instantons are exchanged to twisted sector fields. Although we have used indirect chains of transitions, it would be interesting to study any direct transition mechanism. This is another feature that twisted fields behave like open strings whose boundary conditions are described by heterotic 5-branes. We might have a fuller brainy description for the orbifold fixed points.

We note that some vacua that satisfy the modified modular invariance condition in the presence of heterotic 5-branes give anomalous spectra. Contrary to the perturbative case, the modular invariance condition is not the sufficient condition for the consistency. We still do not know fully sufficient condition for anomaly cancellation in the presence of 5-branes. This is due to lack of the full understanding on the twisted fields; We do not know how they originate, e.g. from branching of a larger representation but at best we obtain them from CFT equation. In the smooth case, Ref. \cite{Choi:2017luj} calculated the modular invariance condition for $E_8\times E_8$ heterotic orbifolds including the 5-brane effects, using the topological vertex and the total instanton number condition is the sufficient condition. 

In this paper, we have focused on the transition mechanism and treated every fixed point is equal. If we have Wilson lines, we may relax the condition. In this way we hope to find a new territory of vacua that might give rise to the Standard Model. Also we may be able to see more nontrivial connections among different vacua, even not being shown connected in this paper. It would also be interesting to study these questions in non-prime orbifolds \cite{CKtoappear}.

\subsection*{Acknowledgements}
We are grateful to Michael Ratz, Soo-Jong Rey for discussions and Angel Uranga for a correspondence.
K.S.C. is grateful to Theory Group of Hokkaido University for hospitality where this work is initiated. He also thanks to Theory Group of University of California, Irvine where this work is finished. K. S. C. is partly supported by grants NRF-2015R1D1A1A01059940 and NRF-2018R1A2B2007163 of National Research Foundation of Korea. T. K. is partly supported by grant MEXT KAKENHI Grant Number JP17H05395.

\appendix

\section{Decomposition of trace}
For the antisymmetric representations of $SU(n)$ we have \cite{Erler:1993zy}
\begin{align}
 \tr_{\bf \frac{n(n-1)}{2}} F^4_{SU(n)} &= (n-8) \tr F^4_{SU(n)} + 3 (\tr F^2_{SU(n)})^2, \quad n \ge 4,\\
 \tr_{\bf \frac{n(n-1)(n-2)}{6}} F^4_{SU(n)} &= \frac12(n^2-17n+54) \tr F^4_{SU(n)} + (3n-12)(\tr F^2_{SU(n)})^2.
\end{align}
We have similar decompositions for the spinorial and antisymmetric of $SO(2n)$
\begin{align}
 \tr_{\bf 2^{n-1}} F^4_{SO(2n)} &= - 2^{n-5} \tr_{\rm v} F^4_{SO(2n)} + 3 \cdot 2^{n-5} (\tr_{\rm v} F^2)^2,\\
 \tr_{\rm v} F^4_{SO(2n)} &= 2 \tr F^4_{SU(n)}.
\end{align}
The absolute normalization is understood in the branching of vector $SO(2n)$ into fundamental of $SU(n), {\bf 2n \to n+\overline n}$. The coefficients for $\tr F^4$ is zero for $SU(n)$ with $n=2,3$ thus the corresponding groups are anomaly free.


\begin{thebibliography}{99}


%\cite{Witten:1995gx}
\bibitem{Witten:1995gx}
  E.~Witten,
  ``Small instantons in string theory,''
  Nucl.\ Phys.\ B {\bf 460} (1996) 541
  doi:10.1016/0550-3213(95)00625-7
  [hep-th/9511030].
  %%CITATION = doi:10.1016/0550-3213(95)00625-7;%%
  
  %\cite{Duff:1996rs}
\bibitem{Duff:1996rs}
  M.~J.~Duff, R.~Minasian and E.~Witten,
  ``Evidence for heterotic / heterotic duality,''
  Nucl.\ Phys.\ B {\bf 465} (1996) 413
  doi:10.1016/0550-3213(96)00059-4
  [hep-th/9601036].
  %%CITATION = doi:10.1016/0550-3213(96)00059-4;%%

%\cite{Ganor:1996mu}
\bibitem{Ganor:1996mu}
  O.~J.~Ganor and A.~Hanany,
  ``Small E(8) instantons and tensionless noncritical strings,''
  Nucl.\ Phys.\ B {\bf 474} (1996) 122
  doi:10.1016/0550-3213(96)00243-X
  [hep-th/9602120].
  %%CITATION = doi:10.1016/0550-3213(96)00243-X;%%
  
  %\cite{Morrison:1996na}
  \bibitem{MV}
%\bibitem{Morrison:1996na} 
  D.~R.~Morrison and C.~Vafa,
  ``Compactifications of F theory on Calabi-Yau threefolds. 1,''
  Nucl.\ Phys.\ B {\bf 473}, 74 (1996)
  doi:10.1016/0550-3213(96)00242-8
  [hep-th/9602114];
  %%CITATION = doi:10.1016/0550-3213(96)00242-8;%%
  %\cite{Morrison:1996pp}

%\bibitem{Morrison:1996pp}
  D.~R.~Morrison and C.~Vafa,
  ``Compactifications of F theory on Calabi-Yau threefolds. 2.,''
  Nucl.\ Phys.\ B {\bf 476} (1996) 437
  doi:10.1016/0550-3213(96)00369-0
  [hep-th/9603161].
  %%CITATION = doi:10.1016/0550-3213(96)00369-0;%%
  
  %\cite{Bershadsky:1996nh}
\bibitem{Bershadsky:1996nh}
  M.~Bershadsky, K.~A.~Intriligator, S.~Kachru, D.~R.~Morrison, V.~Sadov and C.~Vafa,
  ``Geometric singularities and enhanced gauge symmetries,''
  Nucl.\ Phys.\ B {\bf 481} (1996) 215
  doi:10.1016/S0550-3213(96)90131-5
  [hep-th/9605200].
  %%CITATION = doi:10.1016/S0550-3213(96)90131-5;%%
  
  %\cite{Haghighat:2014pva}
\bibitem{Haghighat:2014pva}
  B.~Haghighat, G.~Lockhart and C.~Vafa,
  ``Fusing E-strings to heterotic strings: E+E?H,''
  Phys.\ Rev.\ D {\bf 90} (2014) no.12,  126012
  doi:10.1103/PhysRevD.90.126012
  [arXiv:1406.0850 [hep-th]].
  %%CITATION = doi:10.1103/PhysRevD.90.126012;%%
  
  %\cite{Kim:2014dza}
\bibitem{Kim:2014dza} 
  J.~Kim, S.~Kim, K.~Lee, J.~Park and C.~Vafa,
  ``Elliptic Genus of E-strings,''
  JHEP {\bf 1709}, 098 (2017)
  doi:10.1007/JHEP09(2017)098
  [arXiv:1411.2324 [hep-th]].
  %%CITATION = doi:10.1007/JHEP09(2017)098;%%
  
  %\cite{Choi:2017vtd}
\bibitem{Choi:2017vtd}
  K.~S.~Choi and S.~J.~Rey,
  ``E(lementary)-Strings in Six-Dimensional Heterotic F-Theory,''
  JHEP {\bf 1709} (2017) 092
  doi:10.1007/JHEP09(2017)092
  [arXiv:1706.05353 [hep-th]].
  %%CITATION = doi:10.1007/JHEP09(2017)092;%%
  
  %\cite{Choi:2017luj}
\bibitem{Choi:2017luj}
  K.~S.~Choi and S.~J.~Rey,
  ``Elliptic Genus, Anomaly Cancellation and Heterotic M-theory,''
  arXiv:1710.07627 [hep-th].
  %%CITATION = ARXIV:1710.07627;%%
 
 %\cite{Dixon:1985jw}
\bibitem{DHVW}
  L.~J.~Dixon, J.~A.~Harvey, C.~Vafa and E.~Witten,
  ``Strings on Orbifolds,''
  Nucl.\ Phys.\ B {\bf 261} (1985) 678.
  doi:10.1016/0550-3213(85)90593-0; 
  %%CITATION = doi:10.1016/0550-3213(85)90593-0;%%

  %\cite{Dixon:1986jc}
%\bibitem{Dixon:1986jc}
  L.~J.~Dixon, J.~A.~Harvey, C.~Vafa and E.~Witten,
  ``Strings on Orbifolds. 2.,''
  Nucl.\ Phys.\ B {\bf 274} (1986) 285.
  doi:10.1016/0550-3213(86)90287-7
  %%CITATION = doi:10.1016/0550-3213(86)90287-7;%% 
  
  
 %\cite{Berkooz:1996iz}
\bibitem{Berkooz:1996iz}
  M.~Berkooz, R.~G.~Leigh, J.~Polchinski, J.~H.~Schwarz, N.~Seiberg and E.~Witten,
  ``Anomalies, dualities, and topology of D = 6 N=1 superstring vacua,''
  Nucl.\ Phys.\ B {\bf 475} (1996) 115
  doi:10.1016/0550-3213(96)00339-2
  [hep-th/9605184].
  %%CITATION = doi:10.1016/0550-3213(96)00339-2;%%
  
  
  %\cite{Erler:1993zy}
\bibitem{Erler:1993zy}
  J.~Erler,
  ``Anomaly cancellation in six-dimensions,''
  J.\ Math.\ Phys.\  {\bf 35} (1994) 1819
  doi:10.1063/1.530885
  [hep-th/9304104].
  %%CITATION = doi:10.1063/1.530885;%%
  
  %\cite{Kachru:1995wm}
\bibitem{Kachru:1995wm}
  S.~Kachru and C.~Vafa,
  ``Exact results for N=2 compactifications of heterotic strings,''
  Nucl.\ Phys.\ B {\bf 450} (1995) 69
  doi:10.1016/0550-3213(95)00307-E
  [hep-th/9505105].
  %%CITATION = doi:10.1016/0550-3213(95)00307-E;%%
  
  %\cite{Aldazabal:1995yw}
\bibitem{Aldazabal:1995yw}
  G.~Aldazabal, A.~Font, L.~E.~Ibanez and F.~Quevedo,
  ``Chains of N=2, D = 4 heterotic type II duals,''
  Nucl.\ Phys.\ B {\bf 461} (1996) 85
  doi:10.1016/0550-3213(95)00654-0
  [hep-th/9510093].
  %%CITATION = doi:10.1016/0550-3213(95)00654-0;%%
 
  
  %\cite{Aldazabal:1997wi}
\bibitem{A}
  G.~Aldazabal, A.~Font, L.~E.~Ibanez, A.~M.~Uranga and G.~Violero,
  ``Nonperturbative heterotic D = 6, D = 4, N=1 orbifold vacua,''
  Nucl.\ Phys.\ B {\bf 519} (1998) 239
  doi:10.1016/S0550-3213(98)00007-8
  [hep-th/9706158].
  %%CITATION = doi:10.1016/S0550-3213(98)00007-8;%%
  
  %\cite{Choi:2002fn}
\bibitem{Choi:2002fn}
  K.~S.~Choi and J.~E.~Kim,
  ``Z(2) orbifold compactification of heterotic string and 6-D SO(14) flavor unification model,''
  Phys.\ Lett.\ B {\bf 552} (2003) 81
  doi:10.1016/S0370-2693(02)03104-0
  [hep-th/0206099].
  %%CITATION = doi:10.1016/S0370-2693(02)03104-0;%%
  
  %\cite{Honecker:2006qz}
\bibitem{Honecker:2006qz}
  G.~Honecker and M.~Trapletti,
  ``Merging Heterotic Orbifolds and K3 Compactifications with Line Bundles,''
  JHEP {\bf 0701} (2007) 051
  doi:10.1088/1126-6708/2007/01/051
  [hep-th/0612030].
  %%CITATION = doi:10.1088/1126-6708/2007/01/051;%
  
  %\cite{Intriligator:1997kq}
\bibitem{Intriligator:1997kq}
  K.~A.~Intriligator,
  ``RG fixed points in six-dimensions via branes at orbifold singularities,''
  Nucl.\ Phys.\ B {\bf 496} (1997) 177
  doi:10.1016/S0550-3213(97)00236-8
  [hep-th/9702038].
  %%CITATION = doi:10.1016/S0550-3213(97)00236-8;%%
 
 
   %\cite{Eguchi:1980jx}
\bibitem{Eguchi:1980jx}
  T.~Eguchi, P.~B.~Gilkey and A.~J.~Hanson,
  ``Gravitation, Gauge Theories and Differential Geometry,''
  Phys.\ Rept.\  {\bf 66} (1980) 213.
  doi:10.1016/0370-1573(80)90130-1
  %%CITATION = doi:10.1016/0370-1573(80)90130-1;%%
  
  %\cite{Candelas:1985en}
\bibitem{Candelas:1985en}
  P.~Candelas, G.~T.~Horowitz, A.~Strominger and E.~Witten,
  ``Vacuum Configurations for Superstrings,''
  Nucl.\ Phys.\ B {\bf 258} (1985) 46.
  doi:10.1016/0550-3213(85)90602-9
  %%CITATION = doi:10.1016/0550-3213(85)90602-9;%%
  
  %\cite{Seiberg:1996vs}
\bibitem{Seiberg:1996vs}
  N.~Seiberg and E.~Witten,
  ``Comments on string dynamics in six-dimensions,''
  Nucl.\ Phys.\ B {\bf 471} (1996) 121
  doi:10.1016/0550-3213(96)00189-7
  [hep-th/9603003].
  %%CITATION = doi:10.1016/0550-3213(96)00189-7;%%

  %\cite{Sagnotti:1992qw}
\bibitem{Sagnotti:1992qw}
  A.~Sagnotti,
  ``A Note on the Green-Schwarz mechanism in open string theories,''
  Phys.\ Lett.\ B {\bf 294} (1992) 196
  doi:10.1016/0370-2693(92)90682-T
  [hep-th/9210127].
  %%CITATION = doi:10.1016/0370-2693(92)90682-T;%%
   

  
  %\cite{Witten:1994tz}
\bibitem{Witten:1994tz}
  E.~Witten,
  ``Sigma models and the ADHM construction of instantons,''
  J.\ Geom.\ Phys.\  {\bf 15} (1995) 215
  doi:10.1016/0393-0440(94)00047-8
  [hep-th/9410052].
  %%CITATION = doi:10.1016/0393-0440(94)00047-8;%%

\bibitem{HW}  
  %\cite{Horava:1995qa}
%\bibitem{Horava:1995qa}
  P.~Horava and E.~Witten,
  ``Heterotic and type I string dynamics from eleven-dimensions,''
  Nucl.\ Phys.\ B {\bf 460} (1996) 506
  doi:10.1016/0550-3213(95)00621-4
  [hep-th/9510209];
  %%CITATION = doi:10.1016/0550-3213(95)00621-4;%%
  
  %\cite{Horava:1996ma}
%\bibitem{Horava:1996ma}
  P.~Horava and E.~Witten,
  ``Eleven-dimensional supergravity on a manifold with boundary,''
  Nucl.\ Phys.\ B {\bf 475} (1996) 94
  doi:10.1016/0550-3213(96)00308-2
  [hep-th/9603142].
  %%CITATION = doi:10.1016/0550-3213(96)00308-2;%%
  %1716 citations counted in INSPIRE as of 16 Oct 2018
  
  %\cite{Nibbelink:2007rd}
\bibitem{Nibbelink:2007rd}
  S.~Groot Nibbelink, M.~Trapletti and M.~Walter,
  ``Resolutions of C**n/Z(n) Orbifolds, their U(1) Bundles, and Applications to String Model Building,''
  JHEP {\bf 0703} (2007) 035
  doi:10.1088/1126-6708/2007/03/035
  [hep-th/0701227].
  
  %\cite{Nibbelink:2007pn}
\bibitem{Nibbelink:2007pn}
  S.~Groot Nibbelink, T.~W.~Ha and M.~Trapletti,
  ``Toric Resolutions of Heterotic Orbifolds,''
  Phys.\ Rev.\ D {\bf 77} (2008) 026002
  doi:10.1103/PhysRevD.77.026002
  [arXiv:0707.1597 [hep-th]].
  %%CITATION = doi:10.1103/PhysRevD.77.026002;%%
  
 %\cite{Choi:2006qh}
\bibitem{Choi:2006qh}
  K.~S.~Choi and J.~E.~Kim,
  ``Quarks and leptons from orbifolded superstring,''
  Lect.\ Notes Phys.\  {\bf 696} (2006) 1.
  doi:10.1007/b11681670
  %%CITATION = doi:10.1007/b11681670;%%
  
%\cite{Ibanez:1987pj}
\bibitem{Ibanez:1987pj}
  L.~E.~Ibanez, J.~Mas, H.~P.~Nilles and F.~Quevedo,
  ``Heterotic Strings in Symmetric and Asymmetric Orbifold Backgrounds,''
  Nucl.\ Phys.\ B {\bf 301} (1988) 157.
  doi:10.1016/0550-3213(88)90166-6
  %%CITATION = doi:10.1016/0550-3213(88)90166-6;%%  
 
 %\cite{Katsuki:1989bf}
\bibitem{Katsuki:1989bf} 
  Y.~Katsuki, Y.~Kawamura, T.~Kobayashi, N.~Ohtsubo, Y.~Ono and K.~Tanioka,
  ``Z(n) Orbifold Models,''
  Nucl.\ Phys.\ B {\bf 341}, 611 (1990).
  doi:10.1016/0550-3213(90)90542-L
  %%CITATION = doi:10.1016/0550-3213(90)90542-L;%%
  
  %\cite{Choi:2004wn}
\bibitem{Choi:2004wn}
  K.~S.~Choi, S.~Groot Nibbelink and M.~Trapletti,
  ``Heterotic SO(32) model building in four dimensions,''
  JHEP {\bf 0412} (2004) 063
  doi:10.1088/1126-6708/2004/12/063
  [hep-th/0410232].
  %%CITATION = doi:10.1088/1126-6708/2004/12/063;%%
 
 \bibitem{Dixon}
  L.~J.~Dixon, PhD Thesis (1986), Princeton University.

%\cite{Choi:2003pqa}
\bibitem{Choi:2003pqa}
  K.~S.~Choi, K.~Hwang and J.~E.~Kim,
  ``Dynkin diagram strategy for orbifolding with Wilson lines,''
  Nucl.\ Phys.\ B {\bf 662} (2003) 476
  doi:10.1016/S0550-3213(03)00308-0
  [hep-th/0304243].
  %%CITATION = doi:10.1016/S0550-3213(03)00308-0;%%
 
  

%\cite{Narain:1985jj}
\bibitem{Narain:1985jj}
  K.~S.~Narain,
  ``New Heterotic String Theories in Uncompactified Dimensions < 10,''
  Phys.\ Lett.\  {\bf 169B} (1986) 41.
  doi:10.1016/0370-2693(86)90682-9;
  %%CITATION = doi:10.1016/0370-2693(86)90682-9;%%
  
 %\cite{Narain:1986am}
%\bibitem{Narain:1986am}
  K.~S.~Narain, M.~H.~Sarmadi and E.~Witten,
  ``A Note on Toroidal Compactification of Heterotic String Theory,''
  Nucl.\ Phys.\ B {\bf 279} (1987) 369.
  doi:10.1016/0550-3213(87)90001-0
  %%CITATION = doi:10.1016/0550-3213(87)90001-0;%%
  
  %\cite{Choi:2006hm}
\bibitem{Choi:2006hm}
  K.~S.~Choi,
  ``Unification in intersecting brane models,''
  Phys.\ Rev.\ D {\bf 74} (2006) 066002
  doi:10.1103/PhysRevD.74.066002
  [hep-th/0603186].
  %%CITATION = doi:10.1103/PhysRevD.74.066002;%%

%\cite{Choi:2006th}
\bibitem{Choi:2006th}
  K.~S.~Choi,
  ``Intersecting brane world from type I compactification,''
  Int.\ J.\ Mod.\ Phys.\ A {\bf 22} (2007) 3169
  doi:10.1142/S0217751X07036786
  [hep-th/0610026].
  %%CITATION = doi:10.1142/S0217751X07036786;%%  

\bibitem{SW}
%\cite{Schellekens:1986yi}
%\bibitem{Schellekens:1986yi}
  A.~N.~Schellekens and N.~P.~Warner,
  ``Anomalies and Modular Invariance in String Theory,''
  Phys.\ Lett.\ B {\bf 177} (1986) 317.
  doi:10.1016/0370-2693(86)90760-4;
  %%CITATION = doi:10.1016/0370-2693(86)90760-4;%%
  
  %\cite{Schellekens:1986yj}
%\bibitem{Schellekens:1986yj}
  A.~N.~Schellekens and N.~P.~Warner,
  ``Anomaly Cancellation and Selfdual Lattices,''
  Phys.\ Lett.\ B {\bf 181} (1986) 339.
  doi:10.1016/0370-2693(86)90059-6.
  %%CITATION = doi:10.1016/0370-2693(86)90059-6;%%
  
  %\cite{Berkooz:1996dw}
\bibitem{Berkooz:1996dw}
  M.~Berkooz and R.~G.~Leigh,
  ``A D = 4 N=1 orbifold of type I strings,''
  Nucl.\ Phys.\ B {\bf 483} (1997) 187
  doi:10.1016/S0550-3213(96)00543-3
  [hep-th/9605049].
  %%CITATION = doi:10.1016/S0550-3213(96)00543-3;%%
  
  %\cite{Gorbatov:2001pw}
\bibitem{Gorbatov:2001pw}
  E.~Gorbatov, V.~S.~Kaplunovsky, J.~Sonnenschein, S.~Theisen and S.~Yankielowicz,
  ``On heterotic orbifolds, M theory and type I-prime brane engineering,''
  JHEP {\bf 0205} (2002) 015
  doi:10.1088/1126-6708/2002/05/015
  [hep-th/0108135].
  %%CITATION = doi:10.1088/1126-6708/2002/05/015;%%
  
  %\cite{Forste:2004ie}
\bibitem{Forste:2004ie}
  S.~Forste, H.~P.~Nilles, P.~K.~S.~Vaudrevange and A.~Wingerter,
  ``Heterotic brane world,''
  Phys.\ Rev.\ D {\bf 70} (2004) 106008
  doi:10.1103/PhysRevD.70.106008
  [hep-th/0406208].
  %%CITATION = doi:10.1103/PhysRevD.70.106008;%%
  
  %\cite{Kobayashi:2004ud}
\bibitem{Kobayashi:2004ud} 
  T.~Kobayashi, S.~Raby and R.~J.~Zhang,
  ``Constructing 5-D orbifold grand unified theories from heterotic strings,''
  Phys.\ Lett.\ B {\bf 593}, 262 (2004)
  doi:10.1016/j.physletb.2004.04.058
  [hep-ph/0403065].
  %%CITATION = doi:10.1016/j.physletb.2004.04.058;%%
  
 \bibitem{CKtoappear}
 Work in progress.

  
  \end{thebibliography}
\end{document}